\documentstyle[preprint,aps,floats,epsf,tighten,eqsecnum]{revtex}

\begin{document}
 
\draft

\preprint{
 \parbox{1.5in}{\leftline{JLAB-THY-97-41}
                \leftline{WM-97-112}
                \leftline{nucl-th/9710009 } }  }

\title{Electromagnetic interactions for the two-body spectator equations}

\author{
 J.~Adam, Jr.$^{1,2}$, J.~W.~Van Orden$^{1,3}$, Franz~Gross$^{1,4}$ }
\address{
$^1$Jefferson Lab,
12000 Jefferson Avenue, Newport News, VA 23606\\
$^2$ Nuclear Physics Institute, Czech Academy of Sciences,
CZ-25068 {\v R}e{\v z} near Prague, Czech Republic \\
$^3$Department of Physics, Old Dominion University, Norfolk,
VA 23529 \\
$^4$ Department of Physics,
College of William and Mary, Williamsburg, VA 23185
}

\date{\today}
\maketitle

\begin{abstract}

This paper presents a new non-associative algebra which is used to (i)
show how the spectator (or Gross) two-body equations and electromagnetic
currents can be formally derived from the Bethe-Salpeter equation and
currents if both are treated to all orders, (ii) obtain explicit
expressions for the Gross two-body electromagnetic currents valid to any
order, and (iii) prove that the currents so derived are exactly gauge
invariant when truncated consistently to {\it any\/} finite order.  In
addition to presenting these new results, this work complements and
extends previous treatments based largely on the analysis of sums of
Feynman diagrams.

\end{abstract}


\narrowtext

\section{Introduction}

The two-body spectator (or Gross) equations were first introduced in
1969 and have been developed in a number of subsequent papers
\cite{GR69}.  The treatment of electromagnetic interactions in
this context has also been studied \cite{GR87}--\cite{deutff}.  However,
all of these previous treatments have been largely based on the analysis of
Feynman diagrams, and the equations have been largely derived from this
diagrammatic analysis.  In this paper we present an algebraic derivation
of the equations which is complementary to previous diagrammatic
derivations. More specifically, we develop a new operator algebra which
involves some non-associative rules for the treatment of products of
singular operators.  Once this operator algebra has been carefully
defined and developed, it provides a powerful tool for the formal
manipulation of the equations and permits a careful and detailed
comparison with the Bethe-Salpeter equations.  It also alows us to
derive several new results which would be difficult to derive using a
purely diagrammatic approach.  In applications the relativistic kernel
for either the Bethe-Salpeter equation or the Gross equation is usually
expanded in a perturbation series, and in this paper we obtain, for the
first time, the form of the electromagnetic current operator for the 
Gross equation which is
valid to all orders in this expansion.  We also show explicitly that the
theory conserves the charge of a bound state, and that gauge invariance
is {\it exactly\/} preserved when the theory is truncated to any finite
order, provided only that the strong kernel and the electromagnetic
current operator are both truncated to the same finite order.

This work is a continuation of recent work \cite{norm} in which the
normalization condition for the three-body vertex function was
derived, and also lays the foundation for extension of recent
developments of the three-body Gross equations by Stadler and Gross
\cite{sg}.  The new algebra developed in this paper will be used to
derive, in this forthcoming paper, the electromagnetic current operator
for the three-body Gross equations \cite{avg98}, and we have developed
the formalism here with an eye to this extension.  Spectator
currents have also been independently discussed by Kvinikhidze and
Blankleider \cite{kb97}.  Their discussion is more limited in scope than
ours (here we develop an operator algebra, discuss the connection with the
Bethe-Salpeter equation, and obtain results to all orders), but the results
they do obtain agree with us (see the discussion in Sec.~III below).

A number of other works deriving the electromagnetic current for
various relativistic equations have appeared recently.  Coester
and Riska have derived the current operator for the
Blankenbecler-Sugar equation \cite{cr94} and Devine and Wallace
\cite{dw93} and Phillips and Wallace \cite{pw96} have discussed the
construction of a current operator for use with a relativistic
version of the equal time equation.  Extension of the new
operator formalism presented here  to these other
equations is being studied.  This effort may clarify a
number of issues still unresolved in these treatments.

This paper begins with a brief review of the Bethe-Salpeter equation and
the corresponding current operator.  In Sec.~III we extend this
discussion to the Gross equation, in both the unsymmetrized form for
nonidentical particles and the symmetrized form appropriate for the
description of identical particles.  In Sec.~IV  we present the final
form for the currents and show that the currents appropriate for
identical and nonidentical particles are equivalent.  We also show
that the exact results in the two formalisms (BS and spectator) are
identical if both are calculated to all orders.  Then, in Sec.~V
we use the normalization conditions proved in a previous paper
\cite{norm} to show that the charge of the bound state is conserved by
both theories.  In Sec.~VI we discuss the results when the
perturbation expansions for the kernel and the current operator are
truncated to a finite order, and show that gauge invariance is still
satisfied.  Finally, conclusions are presented in Sec.~VII.

\section{Two-body Bethe-Salpeter Equation}

In this section we review the Bethe-Salpeter formalism.  Our results
are not new, but the brief systematic development given here is needed
both as an introduction to what will follow, and as a description of
the formalism to which the spectator results will be compared.  To
prepare the way, we develop the subject using a conventional operator
formalism.  The need for non-associative operators will not appear
until the next section.

\begin{figure}
\epsfxsize=5in
\centerline{\epsffile{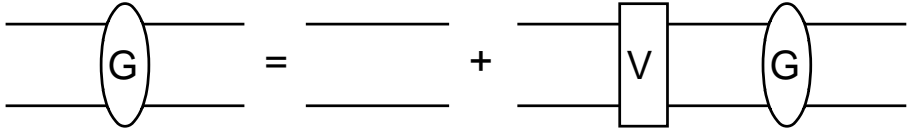}}
\caption{Diagrammatic representation of the integral equation for the
four-point propagator.} \label{four-point}
\end{figure}

The operator form of the equation for the four-point
propagator as  represented  in Fig.~\ref{four-point} is
\begin{eqnarray}
{\cal G}=&&G_{\rm BS}-G_{\rm BS} V {\cal G} \label{g_bethe_salpeter}\\
        =&&G_{\rm BS}-{\cal G} V G_{\rm BS} \, ,
\end{eqnarray}
where the free two-body propagator $G_{\rm BS}=-iG_1G_2$ is defined in
terms of the
single-particle propagators $G_i$ and $V$ is the two-body
Bethe-Salpeter kernel.

The usual momentum-space forms of these expressions can
be obtained by introducing the virtual momentum space states defined such that
\begin{eqnarray}
\left< x\right|\left. p\right>&=&e^{i\bbox{p}\cdot\bbox{x}-ip^0t}\, ,\\
\left< p'\right|\left. p\right>&=&(2\pi)^4\delta^4(p'-p)
\end{eqnarray}
and
\begin{equation}
\int\frac{d^4p}{(2\pi)^4}\left| p\right>\left< p\right|=1\,  .
\end{equation}
The operators are defined such that the momentum matrix elements for the
one-body propagators are
\begin{equation}
\left<p'_i|G_i|p_i\right>=G_i(p_i)(2\pi)^4\delta^4(p'_i-p_i)
\, ,\label{propme}
\end{equation}
the interaction kernel is
\begin{equation}
\left<p'_1p'_2|V|p_1p_2\right>=V(p',p;P)(2\pi)^4\delta^4(P'-P)\, ,
\end{equation}
and the interacting two-body propagator is
\begin{equation}
\left<p'_1p'_2|{\cal G}|p_1p_2\right>={\cal G}(p',p;P)(2\pi)^4
\delta^4(P'-P) \, , \label{propdef}
\end{equation}
where $P=p_1+p_2$ and $P'=p'_1+p'_2$ are the total momenta in the initial and
final states, and $p=\frac{1}{2}(p_1-p_2)$ and $p'=\frac{1}{2}(p'_1-p'_2)$ are
the corresponding relative momenta.

\begin{figure}
\centerline{\epsfxsize=5in\epsffile{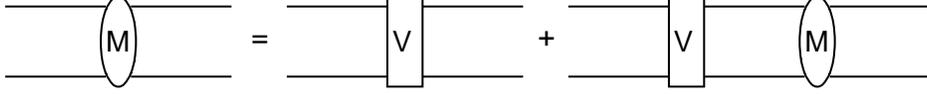}}
\caption{Diagrammatic representation the Bethe-Salpeter equation for the
two-body scattering matrix.}\label{m_bs}
\end{figure}

The two-body propagator can also be written
\begin{equation}
{\cal G}=G_{\rm BS}-G_{\rm BS}{\cal M}G_{\rm BS} \label{G_BSresolved}
\end{equation}
where
\begin{equation}
{\cal M}=V-VG_{\rm BS}{\cal M}=V-{\cal M}G_{\rm BS}V \label{BSM}
\end{equation}
is the two-body scattering matrix. The Bethe-Salpeter equation for the
scattering matrix (\ref{BSM}) is represented by the Feynman diagrams of
Fig.~\ref{m_bs}.
%
%
Equation (\ref{g_bethe_salpeter}) can be written as
\begin{equation}
\left(  G^{-1}_{\rm BS}+V\right) {\cal G}=1
\end{equation}
which implies that the solution for the inverse propagator is
\begin{equation}
{\cal G}^{-1}=G^{-1}_{\rm BS}+V\, . \label{inverg}
\end{equation}

The equation for the Bethe-Salpeter bound-state vertex function is
\begin{equation}
\left|\Gamma\right> = -VG_{\rm BS}\left|\Gamma\right>\, ,
\end{equation}
which can be written
\begin{equation}
0=\left( 1+VG_{\rm BS}\right)\left|\Gamma\right>=\left(
G_{\rm BS}^{-1}+V\right)G_{\rm BS}\left|\Gamma\right>\, .
\end{equation}
Using (\ref{inverg}) this can be written
\begin{equation}
{\cal G}^{-1}\left|\psi\right>=0 \, ,
\end{equation}
where the Bethe-Salpeter bound-state wave function is defined as
\begin{equation}
\left|\psi\right>=G_{\rm BS}\left|\Gamma\right> \, .
\end{equation}

The scattering states are defined in terms of physical, on-shell states
with the normalization
\begin{equation}
\left< x\right|\left. \bbox{p}\right>=e^{i\bbox{p}\cdot\bbox{x}-iE_pt}
\end{equation}
where $E_p=\sqrt{\bbox{p}^2+m^2}$. To include spin, we define the
asymptotic single-particle plane wave momentum state as
\begin{equation}
\left|\bbox{p},s\right>=\left\{
\begin{array}{ll}
u(\bbox{p},s)\left|\bbox{p}\right> & \quad{\rm for\ spin}=\frac{1}{2} \\
\left|\bbox{p}\right> &\quad {\rm for\ spin}=0
\end{array}
\right.
\end{equation}
The final state Bethe-Salpeter scattering wave function with incoming
spherical wave boundary conditions is then
\begin{equation}
\left< \psi^{(-)}\right|=\left< \bbox{p}_1,s_1;\bbox{p}_2,s_2\right|
\left( 1-{\cal M}G_{\rm BS} \right)\,  .
\end{equation}
Using this
\widetext
\begin{eqnarray}
\left< \psi^{(-)}\right|{\cal G}^{-1}&=&\left< \bbox{p}_1,s_1;
\bbox{p}_2,s_2\right|
\left(1-{\cal M}G_{\rm BS} \right)\left(G^{-1}_{\rm BS}+V\right)
\nonumber\\
&=&\left< \bbox{p}_1,s_1;\bbox{p}_2,s_2\right|\left(G^{-1}_{\rm BS}- {\cal
M}+V-{\cal M}G_{\rm BS}V\right) =0 \, ,
\end{eqnarray}
\narrowtext
where (\ref{BSM}) and $\left<\bbox{p}_1,s_1;\bbox{p}_2,s_2\right|
G^{-1}_{\rm BS}=0$   have been used in the last
step. Similarly, the initial state scattering wave function with
outgoing spherical wave boundary conditions
\begin{equation}
\left| \psi^{(+)}\right>=\left(1-G_{\rm BS}{\cal M}\right)\left|
\bbox{p}_1,s_1;\bbox{p}_2,s_2\right>
\end{equation}
satisfies the wave equation
\begin{figure}
\epsfxsize=5in
\centerline{\epsffile{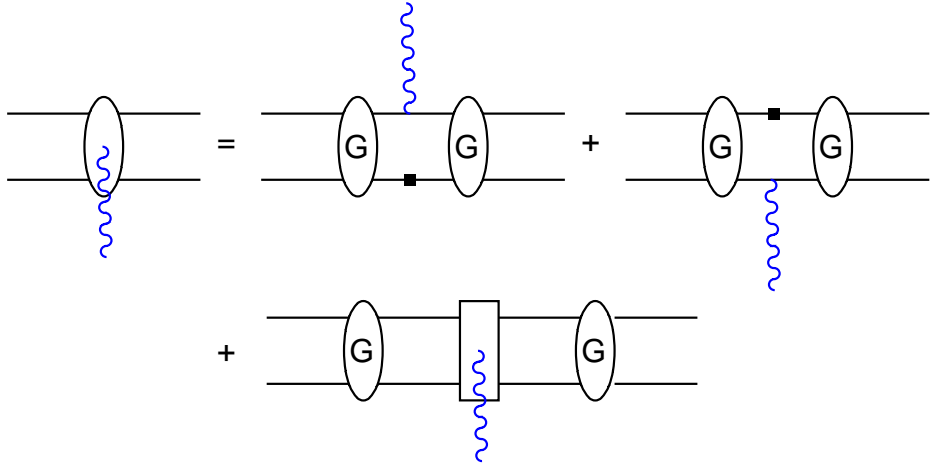}}
\caption{Feynman diagrams representing the five-point
propagator. Inverse one-body propagators are represented by the small,
solid, square boxes inserted on the propagator lines.
}\label{five-point}
\end{figure}
%
\begin{equation}
{\cal G}^{-1}\left| \psi^{(+)}\right>=0\, .
\end{equation}
So the two-body Bethe-Salpeter wave functions for both bound and
scattering  states satisfy the equation
\begin{equation}
{\cal G}^{-1}\left|\psi\right>=\left<\psi\right|{\cal G}^{-1}=0 \, .
\label{wave_eqn_bs}
\end{equation}

The two body current can be obtained from the five-point function
describing the interaction of a photon (with the photon leg amputated)
with the interacting two-body system.  This is represented by the
diagrams of  Fig.~\ref{five-point}, and corresponds to the operator
equation
\begin{equation}
{\cal G}^\mu=-{\cal G}\left( iJ^\mu_1G^{-1}_2+iJ^\mu_2G^{-1}_1+
J^\mu_{\rm ex}
\right) {\cal G} \label{g_muBS}
\end{equation}
where the inverse one-body propagators are introduced to allow for the
factorization in terms of interacting four-point propagators. The
inverse one-body propagators are represented by the square boxes
inserted on the propagator lines in Fig.~\ref{five-point}.

In order to demonstrate that the current
\begin{equation}
J^\mu =iJ^\mu_1G^{-1}_2+iJ^\mu_2G^{-1}_1+J^\mu_{\rm ex}
= J^\mu_{\rm IA}+J^\mu_{\rm ex}
\end{equation}
is conserved, we must introduce the one- and two-body
Ward-Takahashi identities in operator form
\begin{equation}
q_\mu J^\mu_i=\left[e_i(q), G^{-1}_i\right] \label{Ward1}
\end{equation}
and
\begin{equation}
q_\mu J^\mu_{\rm ex}=[e_1(q)+e_2(q),V] \label{Ward2}
\end{equation}
where $e_i(q)$ is the product of the charge $e_i$ (which might be
an operator in isospin space) and a  four-momentum shift
operator defined such that
\begin{equation}
\left<p'_i| e_i(q)|p_i\right>=e_i(2\pi)^4\delta^4(p'_i-p_i-q) \, .
\end{equation}

Using the one- and two-body Ward-Takahashi identities give the
following relation
\widetext
\begin{eqnarray}
q_\mu J^\mu &=& i\left[ e_1(q),G^{-1}_1\right] G^{-1}_2
+i\left[ e_2(q),G^{-1}_2\right] G^{-1}_1+\left[ e_1(q)+e_2(q),V\right]
\nonumber \\
&=&\left[ e_1(q)+e_2(q),G^{-1}_{\rm BS}+V\right]=\left[ e_1(q)+e_2(q),
{\cal G}^{-1}\right]\, .
\end{eqnarray}
\narrowtext
This along with (\ref{wave_eqn_bs}) implies that the two-body current
is conserved
\begin{equation}
q_\mu \left<\psi | J^\mu |\psi \right>=0\, .
\end{equation}

For identical particles, the Bethe-Salpeter
equation can be  rewritten in an explicitly symmetrized form
\begin{equation}
M=\overline{V}-\overline{V}G_{\rm BS}M=\overline{V}-MG_{\rm BS}\overline{V}
\label{BSMSym}
\end{equation}
where $\overline{V}={\cal A}_2V$, $M={\cal A}_2{\cal M}$ and
${\cal A}_2=\frac{1}{2}\left( 1+\zeta{\cal P}_{12}\right)$ is
the two-body symmetrization operator.  (Note that Roman letters (e.g.
$M$) are used for symmetrized quantities and script letters (e.g.
${\cal M}$) for unsymmetrized quantities, as in Ref.~\cite{norm}).  The
corresponding four-point propagator is
\begin{equation}
G={\cal A}_2G_{\rm BS}-G_{\rm BS} \overline{V}  G={\cal A}_2G_{\rm
BS}-G_{\rm BS} M
G_{\rm BS}
\end{equation}
where $G={\cal A}_2{\cal G}$.  The five-point function is also
symmetrized in a similar fashion
\widetext
\begin{eqnarray}
G^\mu&=&- G\left( iJ^\mu_1G^{-1}_2+iJ^\mu_2G^{-1}_1+
\overline{J}^\mu_{\rm ex}\right)  G
\nonumber\\
&=&-{\cal A}_2\left(G_{\rm BS}-G_{\rm BS}MG_{\rm BS}\right)
\left( iJ^\mu_1G^{-1}_2+iJ^\mu_2G^{-1}_1+\overline{J}^\mu_{\rm ex}
\right) \left(G_{\rm BS}-G_{\rm BS}MG_{\rm BS}\right) \label{GmuBS}
\end{eqnarray}
\narrowtext
where $\overline{J}^\mu_{\rm ex}={\cal A}_2 J^\mu_{\rm ex}$ and
satisfies the Ward-Takahashi identity
\begin{equation}
q_\mu \overline{J}^\mu_{\rm ex}=[e_1(q)+e_2(q),\overline{V}] .
\label{Ward3}
\end{equation}
The proof of current conservation follows in exactly the same way as
for the unsymmetrized case.


\section{The Two-Body Gross Equation}

In order to extend this discussion to the spectator or Gross equation,
it is useful to examine the connection of the Gross equation to the
Bethe-Salpeter equation. This is done most easily for the case of
nonidentical particles. Identical particles will be discussed later.


\subsection{Two-Body Equations for Distinguishable Particles}

\begin{figure}
\epsfxsize=1.5in
\centerline{\epsffile{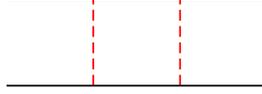}}
\caption{The box diagram.}\label{boxdiagram}
\end{figure}

\begin{figure}
\epsfxsize=2.5in
\centerline{\epsffile{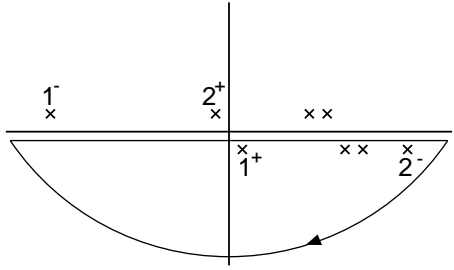}}
\caption{Location of the 8 propagator poles in the integrand of the
box diagram in the complex $p_0$ plane (where $p_0$ is the relative
energy of the two internal particles).}
\label{boxcontour}
\end{figure}

In order to introduce the singular operators needed for our
discussion and to derive their non-associative operator algebra, we
first review the procedure used to motivate the rearrangement of the
multiple scattering series which leads to the Gross equation.  This is
illustrated by considering the second-order box diagram of
Fig.~\ref{boxdiagram} which represents the interaction of two particles
through the exchange of two light bosons.  We assume the two particles
to be of different masses, with the heavier mass associated with
particle 1. The location of the 8 poles in the energy loop integration
is shown in Fig.~\ref{boxcontour}. Here the positive and negative
energy poles of interacting particles 1 and 2 are labeled
$1^{\pm}$ and
$2^{\pm}$ and the poles in the propagators of the  exchanged bosons are
unlabeled.  For low energies the loop integral will be  dominated by
the the two poles $1^+$ and
$2^+$, which lie close to each other (and pinch above the scattering
threshold).  If the contour of integration is closed in the lower
half-plane the result is dominated by the contribution from $1^+$, the
positive energy pole for particle 1.
This suggests that it may be reasonable to separate the contour
integration into two contributions, one containing only the
contribution from the positive energy pole for particle 1 and one
containing contributions from all of the remaining poles within the
contour.

\begin{figure}
\epsfxsize=5in
\centerline{\epsffile{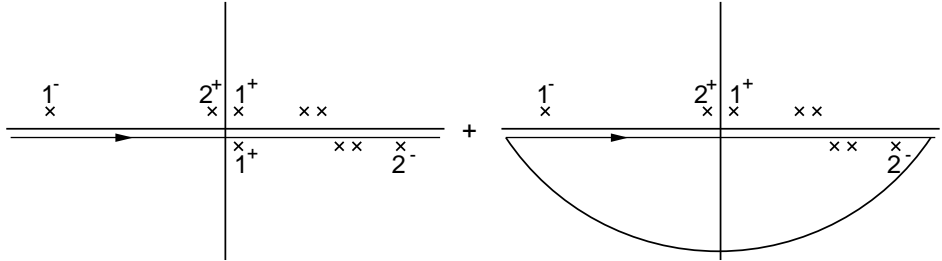}}
\caption{The singularities of the two contributions to the box diagram
resulting from the decomposition of $G_1$ into ${\cal Q}_1$ (left
panel) and $\Delta G_1$ (right panel).  The role of the additional
singularity $1^+$ in the upper half plane in the left panel is to
pinch the contour.  Mathematically this puts particle 1 on-shell.
}\label{separatecontours}
\end{figure}

This separation into two terms is best illustrated by considering the
Dirac propagator for a single particle
\widetext
\begin{eqnarray}
G&=&\int\frac{d^4p}{(2\pi)^4}\left|p\right>\frac{1}{m-\not\!p -i\epsilon}
\left< p\right| \nonumber\\
&=&\int\frac{d^4p}{(2\pi)^4}\left|p\right>\frac{m}{E_p}\left[
\frac{\Lambda^+(\bbox{p})}{E_p-p^0-i\epsilon}+
\frac{\Lambda^-(-\bbox{p})}{E_p+p^0-i\epsilon}\right] \left< p\right|
\label{eq1}\, ,
\end{eqnarray}
\narrowtext
where
\begin{equation}
\Lambda^\pm(\bbox{p}) = {m\pm (E_p\gamma^0-\bbox{p}^i\gamma^i)\over2m}
\end{equation}
are the positive and negative energy projection operators.
If we subtract the the Dirac conjugate
\begin{eqnarray}
{\Lambda^+(\bbox{p})\over E_p-p^0+i\epsilon} \nonumber
\end{eqnarray}
from the first term on the right hand side of (\ref{eq1}) and add it
to the second term we obtain
\widetext
\begin{eqnarray}
G&=&\int\frac{d^4p}{(2\pi)^4}\left|p\right>\left[
\frac{m}{E_p}\frac{2i\epsilon\Lambda^+(\bbox{p})}{\left(E_p-p^0\right)^2+
\epsilon^2}
+\frac{\not\!p+m}{\left(E_p-p^0+i\epsilon\right)
\left(E_p+p^0-i\epsilon\right)}\right] \left< p\right| \label{eq2}\, .
\end{eqnarray}
\narrowtext
The first and second terms on the right hand side of this equation
are represented by the left and right hand diagrams in
Fig.~\ref{separatecontours}, respectively. The first term contains a
new pole which is the conjugate to $1^+$, lies just above the real
axis, and pinches the pole at $1^+$ when the limit $\epsilon\to0$ is
taken.  As we will see below, this automatically selects the
positive energy pole for particle 1.  The second term is a difference
propagator corresponding to the second diagram in Fig.
\ref{separatecontours}.  It is the same as the original
propagator but with pole $1^+$ moved above the real axis.  If we now
define
%
%
\begin{equation}
{\cal Q}=\int\frac{d^4p}{(2\pi)^4}\left|p\right>
\frac{N}{2E_p}\frac{2\epsilon Q(\bbox{p})}{\left(E_p-p^0\right)^2+\epsilon^2}
\left< p\right| \label{calQdef}
\end{equation}
and
\begin{equation}
\Delta G=\int\frac{d^4p}{(2\pi)^4}\left|p\right>
\frac{L(p)}{\left(E_p-p^0+i\epsilon\right)
\left(E_p+p^0-i\epsilon\right)}\left< p\right| \, ,
\end{equation}
where
\begin{equation}
N=\left\{
\begin{array}{ll}
2m & {\rm for\ spin}=\frac{1}{2} \\
1 &  {\rm for\ spin}=0\, ,
\end{array}
\right.
\end{equation}
\begin{equation}
Q(\bbox{p})=\left\{
\begin{array}{ll}
\Lambda^+(\bbox{p}) & {\rm for\ spin}=\frac{1}{2} \\
1 &  {\rm for\ spin}=0 \, ,
\end{array}
\right.
\end{equation}
and
\begin{equation}
L(p)=\left\{
\begin{array}{ll}
\not\!p+m & {\rm for\ spin}=\frac{1}{2} \\
1 &  {\rm for\ spin}=0\, ,
\end{array}
\right.
\end{equation}
we see that the propagator for particle $i$ has been separated into
two pieces
\begin{equation}
G_i=i{\cal Q}_i+\Delta G_i \label{G_1sep}\, .\label{sepofprop}
\end{equation}
Furthermore, using contour integration it is easy to show that
\widetext
\begin{eqnarray}
\lim_{\epsilon\rightarrow 0}{\cal Q}=&&\lim_{\epsilon\rightarrow
0}\int^\infty_{-\infty}\frac{dp^0}{2\pi}\int\frac{d^3p}{(2\pi)^3}
\left|p\right>
\frac{N}{2E_p}\frac{2\epsilon Q(\bbox{p})}{\left(E_p-p^0\right)^2+
\epsilon^2} \left< p\right|\nonumber\\
=&&\int\frac{d^3p}{(2\pi)^3}\frac{N}{2E_p}\left|\bbox{p}\right>
Q(\bbox{p}) \left< \bbox{p}\right| =
\sum_s\int\frac{d^3p}{(2\pi)^3}
\frac{N}{2E_p}\left|\bbox{p},s\right> \left<\bbox{p},s\right|\, .
\label{QdefLim}
\end{eqnarray}
\narrowtext
This shows that ${\cal Q}$ acts to place the propagating particle on
mass  shell and contains the projection operator $Q=Q^2$ on to positive
energy spinor  states, where appropriate.  Be warned that
Refs.~\cite{norm} and \cite{sg} did {\it not\/} make the distinction
between ${\cal Q}$ and $Q$ being made in this paper and that their
${\cal Q}$ is the same as our $Q$.  However, because of the conventions
(\ref{Qconvention}) to be introduced below (which were implicit in
Refs.~\cite{norm} and \cite{sg}), this difference does not affect
the conclusions previously reached in these papers and our results
are consistent with these earlier references.

While the introduction of the operator ${\cal Q}$ may seem
straightforward, it is a singular operator and great care must be
taken when using it.  In particular, like the familiar delta function,
its square is not defined.  Later, we will be faced with the problem of
how to treat quantities which naively appear to be products of
singular operators, or a vanishing operator times a singular operator, and
we will introduce a non-associative algebra for treating these products.
Until then, the analysis is straightforward.

Using (\ref{G_1sep}), the Bethe-Salpeter equation (\ref{BSM}) for the
t-matrix can now be formally separated into a  pair  of coupled
equations.  The first of these is
\begin{equation}
{\cal M}=U-U{\cal Q}_1G_2{\cal M}=U-U{\cal Q}_1g_1{\cal M}\, ,
\label{quasiM}
\end{equation}
or alternately
\begin{equation}
{\cal M}=U-{\cal M}{\cal Q}_1g_1U \, , \label{quasiMr}
\end{equation}
where $U$ is called the quasipotential.   The second equation relates
the quasipotential to the BS kernel $V$.  This equation is derived by
requiring that the scattering matrix ${\cal M}$ as given by
(\ref{quasiM}) be identical to that of (\ref{BSM}). The resulting
equation for the quasipotential is then
\widetext
\begin{equation}
U=V-V(-i\Delta G_1G_2)U=V-
V\Delta g_1U=V-U\Delta g_1V \, .
\label{quasipotential}
\end{equation}
\narrowtext
Note that we use the notation
\begin{eqnarray}
g_1=&&G_2  \nonumber\\
\Delta g_1=&& -i\Delta G_1 G_2 \, ,\label{def}
\end{eqnarray}
where the propagator with particle 1 on  shell is $g_1=G_2$.  [We find
it convenient to label the two-body propagator by the {\it on-shell\/}
particle and to distribute the singular factor of ${\cal Q}_1$ which
accompanies it to other parts of the equation (as discussed below).  We
have therefore introduced the lower case notation (i.e. $g_1$) to
distinguish the off-shell part of the two-body propagator from the one
body propagator
$G_2$.]

\begin{figure}
\centerline{\epsfxsize=5in\epsffile{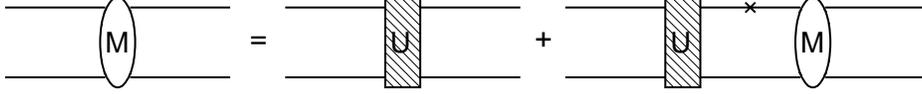}}
\caption{Feynman diagrams representing the Gross equation for the
two-body  scattering matrix. The cross on a propagator line designates
that that  propagator has been placed on its positive energy mass
shell.}
\label{m_gross}
\end{figure}

\begin{figure}
\centerline{\epsfxsize=5in\epsffile{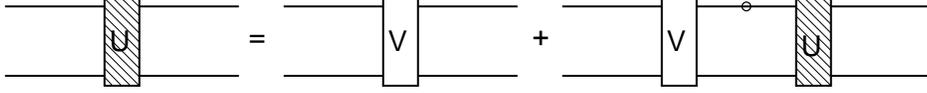}}
\caption{Feynman diagrams representing the quasipotential equation.
The open  circle on a propagator line represents the difference
propagator.}\label{u_gross}
\end{figure}

 The pair of equations (\ref{quasiM}) and (\ref{quasipotential}), as
represented in Figs.~\ref{m_gross} and Fig.~\ref{u_gross}, constitute a
resummation of the multiple scattering series represented by
(\ref{BSM}) and are exactly equivalent to it by construction. The
constrained propagator ${\cal Q}_1g_1$ in (\ref{quasiM}) limits the
phase space available to particle 1 to the positive energy mass shell.
Contributions from the remainder of phase space for particle 1 are
included in the quasipotential (\ref{quasipotential}) through the
difference propagator
$\Delta g_1$.

Using (\ref{QdefLim}) in (\ref{quasiM}) gives
\widetext
\begin{eqnarray}
{\cal M}&=&U-U\int\frac{d^3p_1}{(2\pi)^3}\frac{N}{2E_{p_1}}
\left|\bbox{p}_1\right>
Q_1(\bbox{p}_1) \left< \bbox{p}_1\right|g_1{\cal M}
\nonumber\\
&=&U-U\sum_{s_1}\int\frac{d^3p_1}{(2\pi)^3}\frac{N}{2E_{p_1}}
\left|\bbox{p}_1,s_1\right> \left< \bbox{p}_1,s_1\right|g_1{\cal M}
\, .\label{quasiM2}
\end{eqnarray}
\narrowtext
Note that the projector ${\cal Q}_1$ has introduced
a sum over all on-shell intermediate states for particle 1. In order to
avoid  the necessity of repeatedly writing the on-shell states and the
associated sum,  we will now introduce a notational convention.
We will use the operator $Q_1$ to denote
the presence of on-shell states acting on adjacent operators. If
$Q_1$ appears between two other operators and therefore acts to both the
left  and right, on-shell states acting to both the left and right are
assumed to be  present. In addition the phase-space integral
\begin{equation}
\sum_{s_1}\int\frac{d^3p_1}{(2\pi)^3}\frac{N}{2E_{p_1}}
\end{equation}
is also assumed to be present. If $Q_1$ appears as the first or last
in a  string of operators and therefore acts to the right or left
respectively,  then  only the corresponding on-shell states acting to the
right or  left are assumed.  In this case no phase-space integral is
assumed. That is,
\widetext
\begin{eqnarray}
{\cal O}'Q_1{\cal O}&\Rightarrow& {\cal
O}'\sum_{s_1}\int\frac{d^3p_1}{(2\pi)^3}\frac{N}{2E_{p_1}}
\left|\bbox{p}_1,s_1\right>
\left< \bbox{p}_1,s_1\right|{\cal O} = {\cal O}' {\cal Q}_1 \,{\cal O}
\nonumber\\
{\cal O}Q_1&\Rightarrow& {\cal O}\left|\bbox{p}_1,s_1\right>
\nonumber\\
Q_1{\cal O}&\Rightarrow&\left< \bbox{p}_1,s_1\right|{\cal O}
\label{Qconvention}
\end{eqnarray}
\narrowtext
where ${\cal O}'$ and ${\cal O}$ represent any {\it nonsingular\/}
operators or string of operators.  One consequence of this convention is
the relation
\begin{equation}
{\cal O}'\,Q_1\,{\cal Q}_1\,{\cal O}= {\cal O}'\,{\cal Q}_1\, Q_1\,{\cal O}
= {\cal O}' \,{\cal Q}_1 {\cal O} \,  ,
\label{Qrelation}
\end{equation}
which follows from the observation that $Q_1
\left|\bbox{p}_1,s_1\right>= \left|\bbox{p}_1,s_1\right>$.   Using the
convention (\ref{Qconvention}), we can rewrite (\ref{quasiM2})  as
\begin{eqnarray}
{\cal M}=&&U-UQ_1g_1{\cal M}\label{quasiM3} \\
=&&U-{\cal M}Q_1g_1U\, .\label{quasiM3r}
\end{eqnarray}
We may also replace the $Q_1$ in Eqs.~(\ref{quasiM3}) and
(\ref{quasiM3r}) by $Q_1^2$; in this case the original
Eq.~(\ref{quasiM}) is recovered {\it either\/} by using the convention
(\ref{Qconvention}) on {\it one\/} of the factors of $Q_i$ and then using
(\ref{Qrelation}), {\it or\/}, alternatively, by first replacing $Q_1^2$ by
$Q_1$ and then using the convention (\ref{Qconvention}).  In
Refs.~\cite{norm} and \cite{sg} the $Q$ used here was denoted by ${\cal
Q}$ (and the conventions (\ref{Qconvention}) and (\ref{Qrelation}) were
implicit), so our results agree with those of these previous papers.

Equation (\ref{quasipotential}) represents a four-dimensional integral
equation that is as difficult to solve as the original four-dimensional
Bethe-Salpeter equation. However, as is shown in more detail below, this
equation is usually approximated by iteration and truncation. Equation
(\ref{quasiM3}) can be solved by noting that the constrained propagator
$Q_1g_1$ requires that the scattering matrix on the right hand side of
this equation has particle 1 constrained on shell. Replacing this using
(\ref{quasiM3r}) gives
\begin{equation}
{\cal M}=U-UQ_1g_1Q_1U+UQ_1g_1{\cal M}Q_1g_1U .
\end{equation}
The fully-off-shell t matrix can therefore be obtained by quadrature
from the t matrix with particle 1 constrained on shell in both initial
and final states. This in turn can be obtained by placing particle 1
on-shell  in the initial and final states in
(\ref{quasiM3}) to give
\begin{eqnarray}
{\cal M}_{11}&&=U_{11}-U_{11}g_1{\cal M}_{11} \nonumber\\
&&=U_{11}-{\cal M}_{11}g_1 U_{11}
\label{quasiM4}
\end{eqnarray}
where ${\cal M}_{11}=Q_1{\cal M}Q_1$ and $U_{11}=Q_1UQ_1$.

In order to define the half-off-shell four-point propagator, we want
to replace all of the propagators for particle 1  in
(\ref{G_BSresolved}) with the on-shell projector ${\cal Q}_1g_1$. This
can be done straightforwardly (i.e.~avoiding the appearance of
undefined factors of ${\cal Q}_1^2$) if the free particle inhomogeneous
term is treated separately. We define
\widetext
\begin{equation}
{\cal G}_{11}=Q_1\left[ iG^{-1}_1\left( {\cal
G}-G_{\rm BS}\right)iG^{-1}_1\right] Q_1+Q_1g_1
=Q_1g_1-g_1{\cal M}_{11}g_1
=Q_1g_1-g_1U_{11}{\cal G}_{11} . \label{g11}
\end{equation}
\narrowtext
where the square brackets indicate that the propagators for particle 1
are first amputated using $G^{-1}_1$ and the result is then placed on
shell.  This equation for ${\cal G}_{11}$ can be written
\begin{equation}
\left( {g_1}^{-1}+U_{11}\right){\cal G}_{11}=Q_1 \, . \label{g11inverse}
\end{equation}
Since the projector $Q_1$ does not have an inverse,  ${\cal G}_{11}$
does not  have an inverse. However, the above expression indicates
that ${\cal G}_{11}$ does have an  inverse 
when acting on the subspace spanned by the physical particle states,
ie.~those projected out by the operator $Q_1$.  The solution of
(\ref{g11inverse}) on this subspace will therefore be written
\begin{equation}
{\cal G}_{11}^{-1}= {g_1}^{-1}+U_{11}\, ,
\end{equation}
where we bear in mind that ${\cal G}_{11}$ is defined only on the
space spanned by the physical states of the first particle.

The bound state vertex function for the Gross equation satisfies the
equation
\begin{equation}
\left| \Gamma_1\right>=-U_{11}g_1\left| \Gamma_1\right>\, .
\end{equation}
This can be rewritten
\begin{equation}
0=\left( 1+U_{11}g_1\right) \left| \Gamma_1\right>
=\left({g_1}^{-1}+U_{11}\right) g_1 \left| \Gamma_1\right> \, ,
\end{equation}
or
\begin{equation}
{\cal G}_{11}^{-1}\left| \psi_1\right>=0\, , \label{boundseq}
\end{equation}
where the Gross wave function is defined as
\begin{equation}
\left| \psi_1\right>=g_1 \left| \Gamma_1\right>\,  .
\end{equation}

The final state Gross scattering wave function with incoming spherical
wave boundary conditions is defined as
\begin{equation}
\left< \psi^{(-)}_1\right|=\left< \bbox{p}_1,s_1;\bbox{p}_2,s_2\right|
\left( 1-{\cal
M}_{11}g_1 \right) .
\end{equation}
Using this
\widetext
\begin{eqnarray}
\left< \psi^{(-)}_1\right|{\cal G}^{-1}_{11}&=&
\left< \bbox{p}_1,s_1;\bbox{p}_2,s_2 \right|
\left( 1-{\cal M}_{11}\,g_1 \right)\left({g_1}^{-1}+U_{11}\right)
\nonumber\\
&=&\left< \bbox{p}_1,s_1;\bbox{p}_2,s_2\right|
\left({g_1}^{-1}-{\cal M}_{11}+U_{11}-
{\cal M}_{11}g_1U_{11}\right) =0
\end{eqnarray}
\narrowtext
where (\ref{quasiM4}) and $\left< \bbox{p}_1,s_1;\bbox{p}_2,s_2\right|
{g_1}^{-1}=0$  have been used in the last step.
Similarly, the initial state scattering wave function with outgoing
spherical wave boundary conditions
\begin{equation}
\left| \psi^{(+)}_1\right>=\left(1-g_1{\cal M}_{11}\right)\left|
 \bbox{p}_1,s_1;\bbox{p}_2,s_2\right>
\end{equation}
satisfies the wave equation
\begin{equation}
{\cal G}^{-1}_{11}\left| \psi^{(+)}_1\right>=0 \, .
\end{equation}
So the two-body Gross wave functions for both bound and scattering
states satisfy the equation
\begin{equation}
{\cal G}^{-1}_{11}\left| \psi_1\right>=\left< \psi_1\right|{\cal
G}^{-1}_{11}=0 \, . \label{g11equations}
\end{equation}

\subsection{Two-Body Currents for Distinguishable Particles}

We now turn to the derivation of the two body current operator.  This
will be obtained from the five-point propagator as in our discussion of
the BS equation.

\begin{figure}
\epsfxsize=1.5in
\centerline{\epsffile{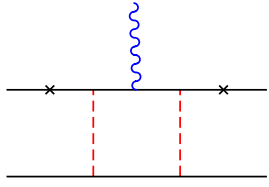}}
\caption{Box diagram with photon insertion.}\label{box_photon}
\end{figure}

\begin{figure}[t]
\epsfxsize=2.5in
\centerline{\epsffile{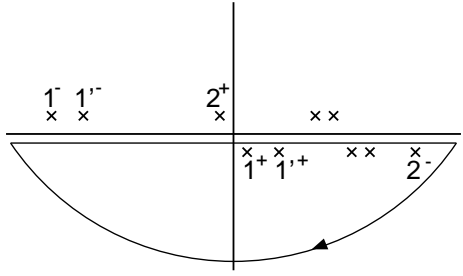}}
\caption{The 10 poles of the box diagram with photon
insertion.}
\label{contour_photon}
\end{figure}

\begin{figure}
\epsfxsize=5in
\centerline{\epsffile{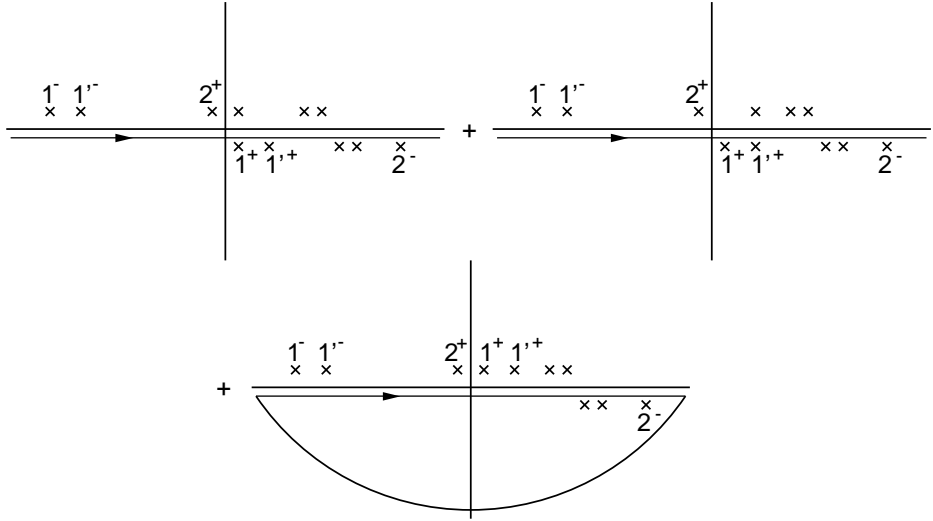}}
\caption{Representation of the three terms resulting from the
decomposition of the propagators of particle 1 in the  presence of the
one-body current insertion.  In the limit $\epsilon\to0$, the pinching
poles in the top two figures insure that particle 1 is on-shell, either
before or after the interaction.  The bottom panel is the contribution
from terms in which particle 1 is off-shell {\it both before and
after\/} the interaction.}
\label{contour_photon2}
\end{figure}

First consider the simple five-point box diagram shown in
Fig.~\ref{box_photon}.
%
%
The location of the 10 poles in the energy loop integral is shown
in Fig.~\ref{contour_photon}.
Since there are now two propagators for particle 1 in the loop, there
are two positive energy poles for particle 1 labeled $1^+$ and
$1^{\prime +}$ corresponding to the two propagators. If the contour is
closed in the  lower  half-plane as shown
in Fig.~\ref{contour_photon}, the contour integral therefore contains
two contributions corresponding to placing particle 1 on shell either
before or after the photon absorption. The separation of  propagators
in the presence of the single nucleon current operator is  then
illustrated by the contour integrals displayed in
Fig.~\ref{contour_photon2}.
For spin 1/2 particles, the contour integral is decomposed into
three terms
\widetext
\begin{eqnarray}
\int_\infty^\infty{dp_0\over 2\pi}&&\int {d^3p\over (2\pi)^3}
\,{\cal O}_f \left({m+\not\!p + {1\over2}\not\!q\over m^2 -
\left(p+{\scriptstyle{1\over2}}q\right)^2-i\epsilon } \right)
J_1^\mu (p,q) \left({m+\not\!p - {1\over2}\not\!q\over m^2 -
\left(p-{\scriptstyle{1\over2}}q\right)^2-i\epsilon }\right)
{\cal O}_i \nonumber\\
=&& i\int {d^3p\over (2\pi)^3}{N\over 2E_+}\,{\cal O}_f\,
\Lambda^+\left( \bbox{p}+{\scriptstyle{1\over2}}\bbox{q}\right)
J_1^\mu (p_+,q) \left({m+\not\!p_+ - {1\over2}\not\!q\over E_-^2 -
\left(E_+ -{\scriptstyle{1\over2}}q_0\right)^2 }\right) {\cal O}_i
\nonumber\\
&&+ \;i\int {d^3p\over (2\pi)^3}{N\over 2E_-}\,{\cal O}_f
\left({m+\not\!p_- + {1\over2}\not\!q\over E_+^2 -
\left(E_- +{\scriptstyle{1\over2}}q_0\right)^2 }\right)
J_1^\mu (p_-,q) \,
\Lambda^+\left( \bbox{p}-{\scriptstyle{1\over2}}\bbox{q}\right)
{\cal O}_i \nonumber\\
&&+ \;{\cal O}_f \Delta G_1 J_1^\mu  \Delta G_1 {\cal O}_i \, ,
\label{eqnx}
\end{eqnarray}
\narrowtext
where ${\cal O}_i$ and ${\cal O}_f$ are operators corresponding to
the particle exchanges which occur before and after the interaction,
$p_\pm=(E_\pm, \bbox{p})$ with
$E_\pm=\sqrt{m^2+(\bbox{p}\pm{1\over2}\bbox{q})^2}$, and the last
term is the remainder of the $dp_0$ integration coming
from all of the poles {\it except\/} $1^+$ and $1^{\prime +}$.  Note
that the singularities which appear in the first two terms after the
integration {\it cancel\/} as $q\to0$ and that therefore the
$i\epsilon$ prescriptions have been dropped from the propagators.
In algebraic form this decomposition can be written
%
\widetext
\begin{equation}
{\cal O}_f\,\left\{ G_1J^\mu_1G_1 \right\}{\cal O}_i \rightarrow
{\cal O}_f \left\{{\cal Q}_1J^\mu_1\Delta G_1+\Delta G_1J^\mu_1{\cal
Q}_1+ \Delta G_1J^\mu_1\Delta G_1\right\}{\cal O}_i \, ,
\label{G_1JG_1sep}
\end{equation}
\narrowtext
where the $\left\{~\right\}$ brackets indicate
that only one loop integration is present even though there are two
operators $G_1$.

Note that the expression (\ref{G_1JG_1sep}) does {\it not\/} contain
the term ${\cal Q}_1J_1^\mu {\cal Q}_1$ which might be expected if the
decomposition (\ref{G_1sep}) were blindly inserted into $G_1J_1^\mu
G_1$.  In order to obtain such a term the contour integration would have
to pick up the two poles at $1^+$ and $1'^+$ {\it simultaneously\/},
which is clearly impossible.  The only sense in which the contour
integration might seem to pick up these two poles simultaneously is
when they coalesce into a single {\it double\/} pole, which can occur
for certain values of the external and internal loop  momenta.
However, even in these special cases the residue theorem
\widetext
\begin{eqnarray}
\int_C dz {f(z)\over (z-z_0)^2}= \int_C dz \left[{f(z_0)\over
(z-z_0)^2} + {f'(z_0)\over (z-z_0)} +R(z) \right] =
2\pi\,i\, f'(z_0) \nonumber
\end{eqnarray}
\narrowtext
shows that the only contribution comes from the single poles which
result from the Laurent expansion of the integrand at the point $z_0$;
there is no contribution from the double singularity itself.  In our
case, when the two poles {\it do\/} coalesce, the combination of
the first two terms on the RHS (\ref{G_1JG_1sep}) gives the correct
result by  producing a derivative term (similar to the $f'(z_0)$ term
in the above example) arising from the cancellation of the singular
parts of each term.

Note that when the current couples to {\it external\/} lines, or when
particle 1 is disconnected from the graph so that there is {\it no loop
integration involved\/}, the term ${\cal Q}_1J_1^\mu {\cal Q}_1$  {\it
will\/} be present.  It vanishes only from internal loops.  

The relationship between various n-point functions as described in
the Bethe-Salpeter formalism and the corresponding quantities for the
Gross equation can always be obtained by a similar procedure. That
is, starting with the Bethe-Salpeter quantity:
\begin{enumerate}
\item Identify all loops contributing to the n-point function.
\item Reduce all redundant products of one-body operators. For
example in (\ref{g_muBS}) use $G_1G^{-1}_1G_1=G_1$.
\item In loops where the photon {\it does not\/} connect to particle 1,
replace the one-body propagators for particle 1 with  (\ref{G_1sep}).
\item In loops where the photon {\it does\/} connect to particle 1,
replace the  quantity $G_1J^\mu_1G_1$ using (\ref{G_1JG_1sep}).
\end{enumerate}

\noindent Careful application of this procedure will always result in a
correct expression for the Gross n-point functions, and is
straightforward when applied to the derivation of the Gross five-point
propagator. However, in the application to three body systems it is
necessary to treat the six- and seven-point functions, and the task of
identifying  all possible configurations of loops in these cases is quite
tedious.  In this case the task is greatly simplified if we develop a few
identities which are equivalent to introducing a non-associative algebra
for the operators which occur in the spectator theory.  These identities
also simplify the discussion of two-body systems, and will therefore be
developed now.  The discussion of the application of these ideas to
three-body systems is postponed for forthcoming paper
\cite{avg98}.

Since ${\cal Q}$ is very singular at the positive energy pole,
considerable care  must be taken in evaluating the product of this
operator with other operators which may also be singular or
vanishing at the pole position. To see this  consider the product
${\cal Q}iG^{-1}{\cal Q}$ for scalar particles. Using
(\ref{calQdef}),
\widetext
\begin{eqnarray}
\lim_{\epsilon\rightarrow 0}{\cal Q}iG^{-1}{\cal Q}
&=&\lim_{\epsilon\rightarrow 0}
\int\frac{d^4p'}{(2\pi)^4}\left|p'\right>
\frac{1}{2E_{p'}}\frac{2\epsilon }{\left(E_{p'}-p'^0\right)^2+\epsilon^2}
\left< p'\right|
iG^{-1}
\nonumber\\
&&\qquad\qquad\times\int\frac{d^4p}{(2\pi)^4}\left|p\right>
\frac{1}{2E_{p}}\frac{2\epsilon }{\left(E_{p}-p^0\right)^2+\epsilon^2}
\left< p\right|
\nonumber\\
&=&\lim_{\epsilon\rightarrow 0}\int\frac{d^4p}{(2\pi)^4}\left|p\right>
\frac{1}{4E_{p}^2}\frac{4\epsilon^2 }{\left[\left(E_{p}-p^0\right)^2
+\epsilon^2\right]^2}i\left(m^2-p^2-i\epsilon\right)
\left< p\right|
\nonumber\\
&=&\lim_{\epsilon\rightarrow 0}\int\frac{d^4p}{(2\pi)^4}\left|p\right>
\frac{1}{4E_{p}^2}\frac{4\epsilon^2i\left(E_p+p^0-i\epsilon\right) }
{\left(E_{p}-p^0-i\epsilon\right)\left(E_{p}-p^0+i\epsilon\right)^2}
\left< p\right|
\nonumber\\
&=&\int\frac{d^3p}{(2\pi)^3}\frac{N}{2E_p}\left|\bbox{p}\right>
\left< \bbox{p}\right| \, ,
\end{eqnarray}
\narrowtext
and a similar result can be obtained for spin-$\frac{1}{2}$ particles.
This implies that
\begin{equation}
i{\cal Q}_i\,G^{-1}_i\,i{\cal Q}_i\rightarrow\, i{\cal Q}_i\,  .
\label{rule1}
\end{equation}
A similar argument leads to the identities
\begin{eqnarray}
i{\cal Q}_i\,G^{-1}_i\,\Delta G_i&=& \Delta G_i \,G^{-1}_i\, i{\cal Q}_i
\rightarrow 0 \, ,\label{rule2}\\
\Delta G_iG^{-1}_i\Delta G_i&\rightarrow&\Delta G_i \, . \label{rule3}
\end{eqnarray}
Note that these identities all refer to products where $G^{-1}_i$ is
inserted between factors of ${\cal Z}^1_i=i{\cal Q}_i$ or
${\cal Z}^2_i=\Delta G_i$, and can
be sumarized by the compact statement
\begin{eqnarray}
{\cal Z}^\ell_i\,G^{-1}_i\,{\cal
Z}^{\ell'}_i\to \delta_{\ell\ell'}\, {\cal Z}^\ell_i\, . \nonumber
\end{eqnarray}
However, repeating the derivation for operators ${\cal O}$ other than
${\cal Z}^\ell_i$  gives new rules:
\begin{eqnarray}
\Delta G_i\,G^{-1}_i\,{\cal O}_i&=&{\cal O}_i\,G^{-1}_i\,\Delta G_i
\rightarrow{\cal O}_i \, ,\label{rule4} \\
i{\cal Q}_i\,G^{-1}_i\,{\cal O}_i&=&{\cal O}_i\,G^{-1}_i\,i{\cal Q}_i
\rightarrow 0 \, .\label{rule5}
\end{eqnarray}
Hence  $\Delta G_i\,G^{-1}_i\to {\bbox 1}$ and $i{\cal Q}_i\,G^{-1}_i
\to 0$ for all operators ${\cal O}_i$ except
${\cal Z}^\ell$.  These strange results can be understood if it is
recognized that the operator algebra is not associative.  When reducing
products of operators the correct procedure is to first look for
combinations of the form ${\cal Z}^\ell_i\;iG^{-1}_i\,{\cal
Z}^{\ell'}_i$ and use rules (\ref{rule1})--(\ref{rule3}) to reduce any
which are present.  After this is done, rules (\ref{rule4}) and
(\ref{rule5}) can be used to further reduce the expression.  Finally
the conventions (\ref{Qconvention}) can be used.
These rules allow us to carry out formal operations on the
operators which would be impossible or meaningless otherwise, and
give us a truly algebraic way to obtain relations.  As an example,
using (\ref{rule1}) permits us to show that the ${\cal G}_{11}$ given
by the simple relation
\begin{equation}
{\cal Q}_1\left[ iG^{-1}_1\,{\cal G}\,
iG^{-1}_1\right] {\cal Q}_1 \rightarrow {\cal G}_{11} \, .
\label{simpleg11}
\end{equation}
is identical to that defined in Eq.~(\ref{g11}).  The latter relation
(\ref{simpleg11}) has the advantage that it provides a more
obvious and intuitive connection to the BS propagator.

Finally, to implement the decomposition (\ref{G_1JG_1sep}) we introduce
the rule
\begin{equation}
{\cal Q}_i\,J^\mu_i\,{\cal Q}_i\rightarrow 0 \, . \label{rule6}
\end{equation}
Using this, we can write
\begin{eqnarray}
G_1J^\mu_1G_1&&=(i{\cal Q}_1+\Delta G_1)\,J^\mu_1\,(i{\cal Q}_1+\Delta
G_1) \nonumber\\
&&\rightarrow i{\cal Q}_1\,J^\mu_1\,\Delta G_1+\Delta
G_1\,J^\mu_1\,{\cal Q}_1+\Delta G_1\,J^\mu_1\,\Delta G_1 \nonumber\\
&&\rightarrow {\cal Q}_1\,J^\mu_1\, G_1+
G_1\,J^\mu_1\,{\cal Q}_1+\Delta G_1\,J^\mu_1\,\Delta G_1 \label{gjg}
\end{eqnarray}
which reproduces (\ref{G_1JG_1sep}). The rule (\ref{rule6}) will
always  produce the correct result {\it when used inside loops\/} and
when used to convert the combination $i{\cal Q}_1\,J^\mu_1\,\Delta G_1$
to $i{\cal Q}_1\,J^\mu_1\,G_1$ in all connected  diagrams.  It agrees
with the current derived diagrammatically by Riska and Gross \cite{GR87}
and with the results obtained in Ref.~\cite{norm}.

However, in a recent paper Kvinikhidze and Blankleider
\cite{kb97} have claimed that the last line in Eq.~(\ref{gjg}) is in
error, and they propose a ``new'' gauged propagator.  Detailed
examination of their result (see the Appendix A) shows that it agrees with
the last line in Eq.~(\ref{gjg}) {\it provided we treat the propagator
$G_1$ as a principle value\/}, neglecting its imaginary part.  But this
is precisely the meaning of equations like (\ref{gjg}).  The role of the
$i\epsilon$ prescription in the propagator [e.g.~as in
$1/(m^2-p^2-i\epsilon)$] is to tell how to evaluate the contour
integral over $p_0$; once this integral has been evaluated and the result
is no longer singular (which is the case for Eq.~(\ref{gjg}) where the
singularities in each of the first two terms on the RHS cancel in the sum)
we are instructed to set $\epsilon$ to zero.  In previously published
work \cite{deutff}--\cite{norm} this was, in fact, done.  Hence the
results of Ref.~\cite{kb97} are identical to ours, and there is no
error in Ref.~\cite{norm}.

We are now ready to use these new rules to reduce the BS five-point
function with particle one on-shell.  This is obtained from ${\cal G}^\mu$
by first amputating the external factors of $G_1$ and then placing
particle one on-shell.  This gives
\widetext
\begin{eqnarray}
{\cal G}_{11}^\mu\equiv {\cal Q}_1\left[iG^{-1}_1{\cal G}^\mu\,
iG^{-1}_1\right] {\cal Q}_1 &=&-\left( {\cal Q}_1G_2-{\cal
Q}_1G_2{\cal M} G_{\rm BS}\right)
\left( iJ^\mu_1G^{-1}_2+iJ^\mu_2G^{-1}_1+J^\mu_{\rm ex}\right)
\nonumber\\
&&\times\left( {\cal Q}_1G_2-G_{\rm BS}{\cal M} {\cal Q}_1G_2 \right)\,  .
\end{eqnarray}
We now want to rearrange this expression so as to identify an
effective current for use when particle one is on-shell.
The basic procedure is to  rewrite the five-point
function so that it has a form similar to  (\ref{g_muBS}), i.e.~a
four-point function {\it with particle one on-shell\/}, followed by a
current, followed  by a four-point function {\it with particle one
on-shell\/}. This is accomplished by rewriting the above expression so
as to include any contributions from the propagation of {\it two\/}
off-shell particles within the effective current operator. To
this end, consider the factor
\begin{eqnarray}
{\cal Q}_1G_2-G_{\rm BS}{\cal M} {\cal Q}_1G_2&=&{\cal Q}_1G_2-\left(
{\cal Q}_1 -i\Delta G_1\right) G_2{\cal M} {\cal Q}_1G_2
\nonumber \\
&=&{\cal Q}_1G_2-{\cal Q}_1 G_2 {\cal M} {\cal Q}_1G_2+i\Delta
G_1G_2\left(U -U{\cal Q}_1G_2 {\cal M}\right) {\cal Q}_1G_2
\nonumber \\
&=&\left( 1-\Delta g_1U \right)\left( {\cal Q}_1g_1-{\cal Q}_1g_1
{\cal M} {\cal Q}_1g_1\right)
\end{eqnarray}
where in the first step the propagator for particle 1 is written in
its separated form, and in the second step the scattering matrix is
iterated using (\ref{quasiM}). Similarly
\begin{equation}
{\cal Q}_1G_2-{\cal Q}_1G_2{\cal M} G_{\rm BS}= \left( {\cal Q}_1g_1-
{\cal Q}_1g_1 {\cal M} {\cal Q}_1g_1\right) \left( 1-U\Delta g_1
\right) .
\end{equation}
This gives
\begin{eqnarray}
{\cal G}_{11}^\mu
&=&- \left( {\cal Q}_1g_1-{\cal Q}_1g_1
{\cal M} {\cal Q}_1g_1\right) \left( 1-U\Delta g_1 \right)
\left( iJ^\mu_1G^{-1}_2+iJ^\mu_2G^{-1}_1+J^\mu_{\rm ex}\right)
\nonumber\\
&&\times\left( 1-\Delta g_1U \right)\left( {\cal Q}_1g_1-{\cal Q}_1
g_1 {\cal M} {\cal Q}_1g_1\right) \nonumber\\
&\rightarrow& - (Q_1g_1 - g_1{\cal M}_{11}g_1)\, J_{11}^\mu\,
(Q_1g_1 - g_1{\cal M}_{11}g_1)= -{\cal G}_{11} J_{11}^\mu
\,{\cal G}_{11}\,  , \label{quasiGmu}
\end{eqnarray}
where we used (\ref{Qrelation}) in the last line and
\begin{eqnarray}
J_{11}^\mu={\cal Q}_1 \left( 1-U\Delta g_1 \right)
\left( iJ^\mu_1G^{-1}_2+iJ^\mu_2G^{-1}_1+J^\mu_{\rm ex}\right)
\left( 1-\Delta g_1U \right){\cal Q}_1 \label{Current1}
\end{eqnarray}
is the effective current for the Gross equation.  This can be broken
into two terms:
\begin{eqnarray}
J^\mu_{\rm IA,eff}=&&{\cal Q}_1 \left( 1-U\Delta g_1 \right)
\left( iJ^\mu_1G^{-1}_2+iJ^\mu_2G^{-1}_1 \right) \left( 1-\Delta g_1U
\right){\cal Q}_1 \nonumber\\
J^\mu_{\rm ex,eff}=&&{\cal Q}_1 \left( 1-U\Delta g_1 \right)
J^\mu_{\rm ex}
\left( 1-\Delta g_1U \right){\cal Q}_1 \, . \label{GrossCurrent1}
\end{eqnarray}
These forms will be used in our discussion of gauge invariance below.

The effective current can be simplified.  Using the rules
(\ref{rule1})--(\ref{rule5}) and (\ref{rule6}) we obtain
\begin{eqnarray}
J^\mu_{11}&=&{\cal Q}_1\biggl[ J^\mu_2-J^\mu_1G_1U-UG_1J^\mu_1
-iU(\Delta G_1J^\mu_1\Delta G_1) G_2 U
\nonumber \\
& & -i U\Delta G_1 (G_2 J^\mu_2G_2) U
+ \left( 1-U\Delta g_1 \right)
J^\mu_{\rm ex}
\left( 1-\Delta g_1U \right)\biggr] {\cal Q}_1 \label{GrossCurrent}
\end{eqnarray}
This form is convenient for calculations.
 
As in the case of the four-point function (\ref{quasiM}),
(\ref{quasiGmu}) is  simply a resummation of the Bethe-Salpeter five
point function (\ref{GmuBS}) with  particle 1 constrained on shell in
the initial and final states. The two  versions of the five-point
function are equivalent by construction. This in  turn guaranties
that the matrix elements of the effective current between  physical
asymptotic states will also be equivalent.
Any matrix element of this effective current is of the form
\begin{eqnarray}
\left< \psi_1\right| J^\mu_{11}\left| \psi_1\right>&=& \left<
\psi_1\right| \biggl[ J^\mu_2-J^\mu_1G_1U-UG_1J^\mu_1
-i U(\Delta G_1J^\mu_1\Delta G_1) G_2 U
\nonumber \\
& & \qquad -i U\Delta G_1 (G_2 J^\mu_2G_2) U
+ \left( 1-U\Delta g_1 \right)
J^\mu_{\rm ex}
\left( 1-\Delta g_1U \right)\biggr] \left| \psi_1\right>\, .
\label{GrossCurrentME}
\end{eqnarray}

We now show that the sum of the currents (\ref{GrossCurrent1}) is
gauge invariant.  If we define $J^\mu_{\rm IA}=
iJ^\mu_1G^{-1}_2+iJ^\mu_2G^{-1}_1$, Eq.~(\ref{Ward1}) can be written
\begin{equation}
q_\mu J^\mu_{\rm IA}=[e_1(q)+e_2(q),G_{\rm BS}^{-1}] \, .
\label{Ward1a}
\end{equation}
Recalling (\ref{Ward2}) the divergences of the two parts of the
effective current become
\begin{eqnarray}
 q_\mu J^\mu_{\rm IA,eff} &=&{\cal Q}_1 \left( 1-U\Delta g_1
\right)[e_1(q)+e_2(q),G_{\rm BS}^{-1}] \left( 1-\Delta g_1U
\right){\cal Q}_1 \nonumber \\
 q_\mu J^\mu_{\rm ex,eff} &=&{\cal Q}_1 \left( 1-U\Delta g_1 \right)
[e_1(q)+e_2(q), V] \left( 1-\Delta g_1U\right){\cal Q}_1 \, .
\end{eqnarray}
Adding these gives
\begin{equation}
 q_\mu J^\mu_{11}= {\cal Q}_1 \left( 1-U\Delta g_1
\right)[e_1(q)+e_2(q), {\cal G}^{-1}] \left( 1-\Delta g_1U
\right){\cal Q}_1 \, . \label{gauge1}
\end{equation}
Next we reduce the factor ${\cal G}^{-1} \left( 1-\Delta g_1U
\right){\cal Q}_1$.  The result we obtain depends on whether $e_1$
or $e_2$ multiplies from the left.  If the factor is $e_1$, use
rules (\ref{rule4}) and (\ref{rule5}) and the equation for the
quasipotential (\ref{quasipotential}) to obtain
\begin{eqnarray}
e_1\,{\cal G}^{-1} \left( 1-\Delta g_1U
\right){\cal Q}_1=&& e_1\, \left(iG_1^{-1}G_2^{-1}\left[1+i(\Delta
G_1)G_2U\right] +V -V
\Delta g_1U\right) {\cal Q}_1\nonumber\\
=&&e_1\,\left(-U+U\right) {\cal Q}_1=0\, .
\end{eqnarray}
\narrowtext
A similar result holds for ${\cal Q}_1\left( 1-U \Delta g_1
\right) {\cal G}^{-1}$,  and we see that  
\begin{equation}
q_\mu J^\mu_{11}\Bigr|_{e_1\;{\rm terms}}=0 \, ,
\end{equation}
{\it independent\/} of the fact that the initial and final states satisfy
Eqs.~(\ref{g11equations}).  Hence Eq.~(\ref{gauge1}) reduces to
\begin{equation}
q_\mu J^\mu_{11}= {\cal Q}_1 \left( 1-U\Delta g_1
\right)[e_2(q), {\cal G}^{-1}] \left( 1-\Delta g_1U
\right){\cal Q}_1 \, . \label{gauge2}
\end{equation}

To further reduce Eq.~(\ref{gauge2}) we first use
Eq.~(\ref{quasipotential}) to simplify terms involving the commutator
$[e_2(q),V]$
\widetext
\begin{eqnarray}
q_\mu J^\mu_{11}=&& {\cal Q}_1\biggl\{ [e_2(q), G_{\rm BS}^{-1}]  - U
\Delta g_1 [e_2(q), G_{\rm BS}^{-1}] - [e_2(q), G_{\rm BS}^{-1}]\Delta
g_1 U + U\Delta g_1 [e_2(q), G_{\rm BS}^{-1}] \Delta g_1 U \nonumber\\
&&  +  [e_2(q),U] + U [e_2(q), \Delta g_1]U
\biggr\} {\cal Q}_1  \nonumber\\
=&& {\cal Q}_1\biggl\{ iG_1^{-1} [e_2(q), G_2^{-1}]  +i U
\Delta G_1 [e_2(q), G_2]U  +  [e_2(q),U] + U [e_2(q), \Delta g_1]U
\biggr\} {\cal Q}_1 \nonumber\\
=&& {\cal Q}_1[ e_2(q) ,G_2^{-1} +U] {\cal Q}_1 = {\cal Q}_1[ e_2(q) ,{\cal
G}^{-1}_{11}] {\cal Q}_1 \, ,
\label{WTspec}
\end{eqnarray}
\narrowtext
where the second equation was obtained using rule (\ref{rule2})
to eliminate some of the terms linear in $U$ and rule (\ref{rule3}) to
simplify the term involving $\Delta g_1 [e_2(q), G_{\rm BS}^{-1}] \Delta
g_1$.  The cancellation of the $U^2$ terms, leading to the third equation,
then follows by substituting for $\Delta g_1$ and noting that $\Delta G_1$
commutes with $e_2$.  Using the
conventions (\ref{Qconvention}), Eqs.~(\ref{WTspec}) and
(\ref{g11equations}) imply that
\begin{equation}
q_\mu \left<\psi_1 | J^\mu_{11}|\psi_1 \right>=0 \, ,
\end{equation}
so the current is conserved.


\subsection{Two-Body Equations for Identical Particles}

We will now extend the derivation of the two-body Gross equations to the
case of identical particles. Although simple arguments can be used to
show that the result will have essentially the same form as those in the
previous section with the substitution of appropriately symmetrized
quantities, we will proceed by considering a completely symmetrical
approach to the construction of the four- and five-point functions. We
will then show that necessary quantities can be reduced to a simpler
non-symmetric form suitable for calculation. In doing so we will
illustrate the approach necessary for constricting the effective
currents for the three-body Gross equation.

Starting with (\ref{BSMSym}) and making a symmetrical replacement of the
one-body propagators in the intermediate state gives
\begin{equation}
M=\overline{V}-\overline{V}\;{\textstyle\frac{1}{2}}\left( {\cal
Q}_1g_1+\Delta g_1+ {\cal Q}_2g_2+\Delta
g_2\right) M
\end{equation}
where $g_2=G_1$ is the propagator for particle 2 on shell, and $\Delta g_2=
-iG_1\Delta G_2$.
This can be rewritten as the pair of equations
\begin{equation}
M=\overline{U}-\overline{U}\,{\textstyle\frac{1}{2}}
\left(Q_1g_1+Q_2g_2\right) M
\end{equation}
and
\begin{equation}
\overline{U}=\overline{V}-\overline{V}\,{\textstyle\frac{1}{2}}
\left(\Delta g_1+\Delta g_2\right) \overline{U} .
\end{equation}
There are now two channels that contribute the Gross equation, one where
particle 1 is on shell and one where particle 2 is on shell. For the purposes
of the following discussion it is convenient to pose the various equations in
terms of a two-dimensional channel space. This can be done by introducing the
vector
\begin{equation}
\bbox{D}=\left(
\begin{array}{c}
1 \\
1
\end{array}
\right)
\end{equation}
and the matrices
\begin{equation}
\bbox{g}_0=\left(
\begin{array}{cc}
g_1 & 0 \\
0 & g_2
\end{array}
\right) ,
\end{equation}
\begin{equation}
\bbox{\Delta g}_0=\left(
\begin{array}{cc}
\Delta g_1 & 0 \\
0 & \Delta g_2
\end{array}
\right)  ,
\end{equation}
\begin{equation}
\bbox{\cal Q}=\left(
\begin{array}{cc}
{\cal Q}_1 & 0 \\
0 & {\cal Q}_2
\end{array}
\right)
\end{equation}
and
\begin{equation}
\bbox{Q}=\left(
\begin{array}{cc}
Q_1 & 0 \\
0 & Q_2
\end{array}
\right) .
\end{equation}
We can now write
\begin{equation}
{\cal Q}_1g_1+\Delta g_1+{\cal Q}_2g_2+\Delta g_2
=\bbox{D}^T\left( \bbox{g}_0 \bbox{\cal Q}+\bbox{\Delta
g}_0\right)
\bbox{D}
\end{equation}
The t-matrix equation is then
\begin{equation}
M=\overline{U}-\overline{U}\bbox{D}^T\,{\textstyle\frac{1}{2}}
\bbox{g}_0\bbox{\cal Q}\bbox{D}\,M
=\overline{U}- \overline{U}\bbox{D}^T\,{\textstyle\frac{1}{2}}
\bbox{g}_0\bbox{Q}\bbox{D}\, M \, ,
\end{equation}
where in the last step the limit $\epsilon \to 0$ was taken,
and the corresponding quasipotential equation is
\begin{equation}
\overline{U}=\overline{V}-\overline{V}\,
{\textstyle\frac{1}{2}}
\bbox{D}^T \bbox{\Delta g}_0 \bbox{D}\,\overline{U} \, .
\label{matrixquasi}
\end{equation}
Note that the factor $1/2$ could be included in the definitions
of $\bbox{g}_0$ and $\bbox{\Delta g}_0$. We have chosen not to
do this since the corresponding factors for the three-body case
cannot be subsumed into the propagators.
A closed form for the half-off-shell t-matrices is given by
\begin{equation}
\bbox{Q}\bbox{D}M\bbox{D}^T\bbox{Q}=\bbox{Q}\bbox{D}\overline{U}
\bbox{D}^T\bbox{ Q}
-\bbox{Q}\bbox{D}\overline{U}\,{\textstyle\frac{1}{2}}
\bbox{D}^T\bbox{g}_0
\bbox{Q}\bbox{D}M\bbox{D}^T \bbox{Q}
\end{equation}
Defining the t-matrix as a two-dimensional matrix in the channel
space
\begin{equation}
\bbox{M}=\bbox{Q}\bbox{D}M\bbox{D}^T\bbox{Q}
\end{equation}
and the quasipotential in the channel space
\begin{equation}
\overline{\bbox{U}}=\bbox{Q}\bbox{D}\overline{U}\bbox{D}^T\bbox{Q}
\, ,
\end{equation}
the matrix form of the t-matrix equation is
\begin{equation}
\bbox{M}=\overline{\bbox{U}}-\overline{\bbox{U}}
\,{\textstyle\frac{1}{2}}
\bbox{g}_0\bbox{M} =
\overline{\bbox{U}}-\bbox{M}\,{\textstyle\frac{1}{2}}
\bbox{g}_0\overline{\bbox{U}}
\label{GrossMSymm}
\end{equation}
The nonlinear form of the t-matrix equation is
\begin{equation}
\bbox{M}=\overline{\bbox{U}}
-\bbox{M}\,{\textstyle\frac{1}{2}}\bbox{g}_0\bbox{M}
-\bbox{M}\,{\textstyle\frac{1}{2}}\bbox{g}_0\overline{\bbox{U}}
\,{\textstyle\frac{1}{2}}\bbox{g}_0\bbox{M} \, .
\end{equation}

Next the half-off-shell t matrix is parameterized in terms of a
contribution from a bound state pole at $P^2=M^2$ and a residual part
\begin{equation}
\bbox{M}=\frac{\left|
\bbox{\Gamma}\right>
\left< \bbox{\Gamma}\right|}{P^2-M^2}+\bbox{R}\,  ,
\end{equation}
where the bound state vertex functions are described by the vector of
vertex functions with particle 1 or particle 2 on shell with
\begin{equation}
\left|\bbox{\Gamma}\right>=\left(
\begin{array}{c}
\left|\Gamma_1\right> \\[0.1in]
\left|\Gamma_2\right>
\end{array}\right) \, .
\end{equation}
Using the usual techniques, this gives the fully
symmetrized two-body Gross equation for the bound state vertex function
\begin{equation}
\left|
\bbox{\Gamma}\right>=-\overline{\bbox{U}}\,{\textstyle\frac{1}{2}}
\bbox{g}_{0}
\left| \bbox{\Gamma}\right> \label{GammaGrossSym}
\end{equation}
with normalization given by
\begin{equation}
1=\left< \bbox{\Gamma}\right|\left(\frac{1}{2}\frac{\partial
\bbox{g}_{0}}{\partial P^2}-\frac{1}{2}\bbox{g}_{0}\frac{\partial
\overline{\bbox{U}}}{\partial P^2}\frac{1}{2}\bbox{g}_{0}\right)  \left|
\bbox{\Gamma}\right> .
\end{equation}

It is convenient to introduce the following definition for the
interacting spectator propagator:
\widetext
\begin{equation}
\bbox{g}=\,{\textstyle\frac{1}{2}}
\bbox{g}_0\bbox{Q}-\,{\textstyle\frac{1}{2}}\bbox{g}_0\bbox{M}
\,{\textstyle\frac{1}{2}} \bbox{g}_0
={\textstyle\frac{1}{2}}\bbox{g}_0\bbox{Q}-{\textstyle\frac{1}{2}}
\bbox{g}_0\overline{\bbox{U}}\bbox{g}
={\textstyle\frac{1}{2}}\bbox{g}_0\bbox{Q}-\bbox{g}
\overline{\bbox{U}}\,{\textstyle\frac{1}{2}}\bbox{g}_0\,  .
\label{symmg}
\end{equation}
\narrowtext
This can be rewritten
\begin{equation}
\left(2\bbox{g}_0^{-1}+\overline{\bbox{U}}\right)\bbox{g}
=\bbox{Q} .
\end{equation}
So the ``inverse'' of the propagator is
\begin{equation}
\bbox{g}^{-1}=2\bbox{g}_0^{-1}+\overline{\bbox{U}} .
\end{equation}
The Gross equation for the bound state vertex function
(\ref{GammaGrossSym}) can therefore be rewritten
\begin{equation}
0=\left(\bbox{1}+\overline{\bbox{U}}{\textstyle\frac{1}{2}}
\bbox{g}_{0}\right)
\left|\bbox{\Gamma}\right>=\left(2\bbox{g}_{0}^{-1}+
\overline{\bbox{U}}\right) \frac{1}{2}\, \bbox{g}_{0}
\left|\bbox{\Gamma}\right>
=\bbox{g}^{-1}\left|\bbox{\psi}\right>
\end{equation}
where the Gross bound state wave function is defined by
\begin{equation}
\left|\bbox{\psi}\right>=\frac{1}{2}\,\bbox{g}_0\left|
\bbox{\Gamma}\right> =\frac{1}{2}\left(
\begin{array}{c}
\left|\psi_1\right> \\[0.1in]
\left|\psi_2\right>
\end{array}\right) \, .
\end{equation}

The final state Gross scattering wave function with incoming
spherical wave boundary conditions is defined to be
\widetext
\begin{equation}
\left<\bbox{\psi}^{(-)}\right|=\left< \bbox{p}_1,s_1;
\bbox{p}_2,s_2\right| {\cal A}_2\,
{\textstyle\frac{1}{2}}\bbox{D}^T\left(\bbox{1}-\bbox{M}
{\textstyle\frac{1}{2}}\bbox{g}_0\right) =\frac{1}{2}\biggl(
\begin{array}{cc}
\left<\psi^{(-)}_1\right|\; &\left<\psi^{(-)}_2\right|
\end{array}
\biggr) \, .
\end{equation}
Using this
\begin{eqnarray}
\left<\bbox{\psi}^{(-)}\right|\bbox{g}^{-1}&=&
\left<\bbox{\psi}^{(-)}\right|=\left<\bbox{p}_1,s_1;
\bbox{p}_2,s_2 \right| {\cal A}_2\,
{\textstyle\frac{1}{2}}\bbox{D}^T\left(\bbox{1}-\bbox{M}
{\textstyle\frac{1}{2}}\bbox{g}_0\right)
\left(2\bbox{g}_0^{-1}+\overline{\bbox{U}}\right)
\nonumber\\
&=&\left<\bbox{p}_1,s_1;\bbox{p}_2,s_2\right|{\cal A}_2\,
{\textstyle\frac{1}{2}}\bbox{D}^T\left(2\bbox{g}_0^{-1}-
\bbox{M}+\overline{\bbox{U}}
-\bbox{M}{\textstyle\frac{1}{2}}\bbox{g}_0
\overline{\bbox{U}}\right)=0
\end{eqnarray}
where (\ref{GrossMSymm}) and
$\left<\bbox{p}_1,s_1;\bbox{p}_2,s_2 \right| G_i^{-1}=0$,
for $i= 1,2$,  have been used in
the last step. Similary, the initial state Gross scattering wave
function with outgoing spherical wave boundary conditions
\begin{equation}
\left|\bbox{\psi}^{(+)}\right>=\left(\bbox{1}
-{\textstyle\frac{1}{2}}\bbox{g}_0\bbox{M}\right)\bbox{D}\,
\frac{1}{2}{\cal A}_2
\left|\bbox{p}_1,s_1;\bbox{p}_2,s_2  \right>=\frac{1}{2}\left(
\begin{array}{c}
\left|\psi^{(+)}_1\right> \\[0.2in]
\left|\psi^{(+)}_2\right>
\end{array}\right)
\end{equation}
\narrowtext
satisfies the wave equation
\begin{equation}
\bbox{g}^{-1}\left|\bbox{\psi}^{(+)}\right>=0 .
\end{equation}
So the two-body Gross wave functions for both bound and scattering
states satisfy the equation
\begin{equation}
\bbox{g}^{-1}\left|\bbox{\psi}\right>=\left<\bbox{\psi}\right|
\bbox{g}^{-1} =0\, . \label{symmwavee}
\end{equation}

\subsection{Two-Body Currents for Identical Particles}

Finally, we turn to the construction of the current for identical
particles.  Following the method previously developed, we obtain the
current from the symmetrized five-point propagator for the Gross equation.
This propagator is  obtained from the symmetrized five-point propagator for
the Bethe-Salpeter equation,  Eq.~(\ref{GmuBS}), by replacing the
two-body propagator, $G_{\rm BS}$, associated with internal loops by the
decomposition
\begin{eqnarray}
G_{\rm BS}\,&&={\textstyle\frac{1}{2}}\left( g_1{\cal Q}_1+\Delta g_1
+g_2{\cal Q}_2+\Delta g_2\right) \nonumber\\
&&\rightarrow {\textstyle\frac{1}{2}} \left(\bbox{g}_0\bbox{\cal Q}
+\bbox{\Delta g}_0\right) \, .
\end{eqnarray}
However, since the impulse term contains only one loop and the exchange
term contains two loops, this substitution leads to a different result for
these two cases.  To illustrate this, consider the two-loop combination
$MG_{\rm BS}\overline{J}^\mu_{\rm ex} G_{\rm BS}M$ which involves the exchange
current. This combination gives
\widetext
\begin{eqnarray}
MG_{\rm BS}\overline{J}^\mu_{\rm ex}G_{\rm BS}M&=&M{\textstyle\frac{1}{2}}
\left( g_1{\cal  Q}_1+\Delta g_1
+g_2{\cal Q}_2+\Delta g_2\right)\overline{J}^\mu_{\rm ex}
\nonumber\\
&&\;\;\times{\textstyle\frac{1}{2}} \left(g_1{\cal Q}_1+\Delta g_1
+g_2{\cal Q}_2+\Delta g_2\right) M
\nonumber\\
&=&M\bbox{D}^T {\textstyle\frac{1}{2}} \left(\bbox{g}_0\bbox{\cal Q}
+\bbox{\Delta g}_0\right)
\bbox{D} \overline{J}^\mu_{\rm ex}
\bbox{D}^T{\textstyle\frac{1}{2}} \left(\bbox{g}_0\bbox{\cal Q}
+\bbox{\Delta g}_0\right)\bbox{D}M
\nonumber\\
&=&M\bbox{D}^T{\textstyle\frac{1}{2}}\left(\bbox{g}_0\bbox{\cal Q}
+\bbox{\Delta g}_0\right)
\overline{\bbox{J}}^\mu_{\rm ex} {\textstyle\frac{1}{2}}
\left(\bbox{g}_0\bbox{\cal Q} +\bbox{\Delta g}_0\right)\bbox{D}M
\end{eqnarray}
where $\overline{\bbox{J}}^\mu_{\rm ex}=\bbox{D}
\overline{J}^\mu_{\rm ex} \bbox{D}^T$. Note that the factors of $1/2$
are the result of the fact that each of the two independent loops
can be closed in two different ways. 

The comparable combination
for the one-body current $G_{\rm BS}J^\mu_{\rm IA}G_{\rm BS}$ contains only
one energy-momentum  loop  that can be closed in either
of two ways so the symmetric separation of the
propagators gives
\begin{eqnarray}
MG_{\rm BS}J^\mu_{\rm IA}G_{\rm BS}M&=&M{\textstyle\frac{1}{2}}
\bigl[\left(g_1{\cal Q}_1+\Delta g_1\right)
J^\mu_{\rm IA}\left(g_1{\cal Q}_1+\Delta g_1\right)
\nonumber\\
&&\;\;+\left(g_2{\cal Q}_2+\Delta g_2\right)
J^\mu_{\rm IA}\left(g_2{\cal Q}_2+\Delta g_2\right)\bigr]M
\nonumber\\
&=&M{\textstyle\frac{1}{2}}\bbox{D}^T\left(\bbox{g}_0\bbox{\cal Q}
+\bbox{\Delta g}_0\right) J^\mu_{\rm IA}\left(\bbox{g}_0\bbox{\cal
Q}+\bbox{\Delta g}_0\right)
\bbox{D}M
\nonumber\\
&=&M{\textstyle\frac{1}{2}}\bbox{D}^T\left(\bbox{g}_0\bbox{\cal Q}
+\bbox{\Delta g}_0\right) 2\bbox{J}^\mu_{\rm IA}
{\textstyle\frac{1}{2}}\left(\bbox{g}_0\bbox{\cal Q} +\bbox{\Delta
g}_0\right)\bbox{D}M
\end{eqnarray}
where $\bbox{J}^\mu_{\rm IA}=J^\mu_{\rm IA}\bbox{1}$.  This argument
shows that the factor $J^\mu_{\rm IA}$ is transformed into $2
\bbox{J}^\mu_{\rm IA}$.  To complete the symmetrization we make the
substitutions
\begin{equation}
G_{\rm BS}\rightarrow {\textstyle\frac{1}{2}}\;\bbox{g}_0\bbox{\cal Q}
\end{equation}
for external two-body propagators,
\begin{equation}
M\rightarrow \bbox{D}M\bbox{D}^T
\end{equation}
for the t matrix, and
\begin{equation}
J^\mu_{\rm IA} +\overline{J}^\mu_{\rm ex}\rightarrow
2\bbox{J}^\mu_{\rm IA} +\overline{\bbox{J}}^\mu_{\rm ex}
\end{equation}
for the current. This transforms (\ref{GmuBS}) into
\begin{eqnarray}
G^\mu\rightarrow &&
-{\cal A}_2\,{\textstyle\frac{1}{2}}\,\bbox{g}_0 \bbox{\cal Q}
\left[\bbox{1}-\bbox{D}M\bbox{D}^T
{\textstyle\frac{1}{2}}\left(\bbox{g}_0\bbox{\cal Q}+ \bbox{\Delta
g}_0\right)\right]
\left(2\bbox{J}^\mu_{\rm IA} +\overline{\bbox{J}}^\mu_{\rm ex}\right)
\nonumber\\
&&\qquad\times\left[\bbox{1}-{\textstyle\frac{1}{2}}\left(\bbox{g}_0
\bbox{\cal Q}+ \bbox{\Delta g}_0\right)
\bbox{D}M\bbox{D}^T\right]{\textstyle\frac{1}{2}}\bbox{g}_0
\bbox{\cal Q} .
\label{GrossGmu1}
\end{eqnarray}

As before, it is convenient to simplify the five-point function by
incorporating any appearence of off-shell two-body propagators
within the effective current operator. To do this, consider the
factor
\begin{eqnarray}
\bbox{1}-{\textstyle\frac{1}{2}}\left(\bbox{g}_0\bbox{\cal
Q}+\bbox{\Delta g}_0\right)\bbox{D}M\bbox{D}^T
&=&\bbox{1}-{\textstyle\frac{1}{2}}\bbox{g}_0\bbox{\cal Q}
\bbox{D}M\bbox{D}^T -{\textstyle\frac{1}{2}}\bbox{\Delta g}_0
\bbox{D}M\bbox{D}^T \nonumber\\
&=&\bbox{1}-{\textstyle\frac{1}{2}}\bbox{g}_0\bbox{\cal Q}
\bbox{D}M\bbox{D}^T -{\textstyle\frac{1}{2}}\bbox{\Delta
g}_0\bbox{D}\overline{U}\bbox{D}^T\left(\bbox{1}
-{\textstyle\frac{1}{2}}\bbox{g}_0\bbox{\cal Q}\bbox{D}M
\bbox{D}^T\right) \nonumber\\
&=&\left(\bbox{1}-{\textstyle\frac{1}{2}}\bbox{\Delta
g}_0\bbox{D}\overline{U}\bbox{D}^T\right)
\left(\bbox{1}-{\textstyle\frac{1}{2}}\bbox{g}_0\bbox{\cal Q}
\bbox{D}M\bbox{D}^T\right) .
\end{eqnarray}
Similarly,
\begin{equation}
\bbox{1}-\bbox{D}M\bbox{D}^T
{\textstyle\frac{1}{2}}\left(\bbox{g}_0\bbox{\cal Q}+
\bbox{\Delta g}_0\right)
=\left(\bbox{1}-\bbox{D}M\bbox{D}^T{\textstyle\frac{1}{2}}
\bbox{g}_0\bbox{\cal Q}\right)
\left(\bbox{1}-\bbox{D}\overline{U}\bbox{D}^T{\textstyle\frac{1}{2}}
\bbox{\Delta g}_0\right)  .
\end{equation}
Equation (\ref{GrossGmu1}) can then be rewritten
\begin{eqnarray}
G^\mu\rightarrow &&
-{\cal A}_2\,{\textstyle\frac{1}{2}}\bbox{g}_0\bbox{\cal Q}
\left(\bbox{1}-\bbox{D}M\bbox{D}^T{\textstyle\frac{1}{2}}
\bbox{g}_0
\bbox{\cal Q}\right)
\left(\bbox{1}-\bbox{D}\overline{U}\bbox{D}^T{\textstyle\frac{1}{2}}
\bbox{\Delta g}_0\right)
\left(2\bbox{J}^\mu_{\rm IA} +\overline{\bbox{J}}^\mu_{\rm ex}\right)
\nonumber\\
&&\qquad\times\left(\bbox{1}-{\textstyle\frac{1}{2}}\bbox{\Delta
g}_0\bbox{D}\overline{U}\bbox{D}^T\right)
\left(\bbox{1}-{\textstyle\frac{1}{2}}\bbox{g}_0\bbox{\cal Q}
\bbox{D}M\bbox{D}^T\right)
{\textstyle\frac{1}{2}}\bbox{g}_0\bbox{\cal Q} .
\end{eqnarray}
Using Eq.~(\ref{symmg}) for the symmetric propagator, $\bbox{g}$,
this becomes
\begin{equation}
\bbox{g}^\mu=-{\cal A}_2 \bbox{g}\bbox{J}^\mu\bbox{g}
\end{equation}
where the matrix current operator $\bbox{J}^\mu$ is given by
\begin{eqnarray}
\bbox{J}^\mu
=&&\bbox{\cal Q}\left(\bbox{1}-\bbox{D}\overline{U}\bbox{D}^T
{\textstyle\frac{1}{2}}
\bbox{\Delta  g}_0\right)
\left(2\bbox{J}^\mu_{\rm IA} +\overline{\bbox{J}}^\mu_{\rm ex}\right)
\left(\bbox{1}-{\textstyle\frac{1}{2}}\bbox{\Delta g}_0\bbox{D}
\overline{U}\bbox{D}^T\right) \bbox{\cal Q}\nonumber\\
=&&\bbox{J}^\mu_{\rm IA,eff}+\bbox{J}^\mu_{\rm ex,eff}\, .
\label{symmcur}
\end{eqnarray}
\narrowtext
We note for future reference that, using the rules
(\ref{rule1}--\ref{rule5}) and (\ref{rule6}), the contributions
from the one-body current can be simplified,
\begin{equation}
\bbox{\cal Q}\bbox{J}^\mu_{\rm IA}\bbox{\cal Q}\rightarrow
\bbox{Q}
\left(
\begin{array}{cc}
J^\mu_2& 0  \\
0 & J^\mu_1  \\
\end{array}
\right)\bbox{Q}  \, , \label{symm1}
\end{equation}
\begin{equation}
\bbox{\cal Q}\bbox{J}^\mu_{\rm IA}\bbox{\Delta g}_{0}
\rightarrow\bbox{Q}\left(
\begin{array}{cc}
J^\mu_1\Delta G_1& 0 \\
0 & J^\mu_2\Delta G_2
\end{array}
\right)\, , \label{symm2}
\end{equation}
\begin{equation}
\bbox{\Delta g}_{0}\bbox{J}^\mu_{\rm IA}\bbox{\cal Q}
\rightarrow\left(
\begin{array}{cc}
\Delta G_1J^\mu_1 & 0 \\
0 & \Delta G_2J^\mu_2
\end{array}
\right)\bbox{Q} \, ,\label{symm3}
\end{equation}
and
\widetext
\begin{equation}
\bbox{\Delta g}_{0}\bbox{J}^\mu_{\rm IA}\bbox{\Delta g}_{0}
\rightarrow -i\left(
\begin{array}{cc}
\Delta G_1 J_1^\mu \Delta G_1 G_2 + G_2J^\mu_2G_2\Delta G_1 & 0 \\
0 & \Delta G_2 J_2^\mu \Delta G_2 G_1 +G_1J^\mu_1G_1\Delta G_2
\end{array}\right) \, . \label{symm4}
\end{equation}

We conclude this section with a discussion of the proof of gauge
invariance for the symmetric current (\ref{symmcur}).  The matrix
form of one body Ward identity, Eq.~(\ref{Ward1}), is
\begin{equation}
q_\mu \bbox{J}^\mu_{\rm IA}=[e_1(q)+e_2(q),G_{\rm BS}^{-1}]\bbox{1}
\, ,
\label{Ward1am}
\end{equation}
\narrowtext
and using the two-body Ward identity, (\ref{Ward2}), together with
the equation for the quasipotential (\ref{matrixquasi})
and rules (\ref{rule1}--\ref{rule4}),  the
four-divergences of the two parts of the effective current are
\widetext
\begin{eqnarray}
 q_\mu \bbox{J}^\mu_{\rm IA,eff} &=&\bbox{\cal Q}\,
2[e_1(q)+e_2(q),G_{\rm BS}^{-1}]\bbox{\cal Q} \nonumber\\
&&- \bbox{\cal Q}\, [e_1(q)+e_2(q),G_{\rm BS}^{-1}]\, 
 \bbox{\Delta g}_0 \bbox{D}\overline{U}\bbox{D}^T \bbox{\cal Q} 
 - \bbox{\cal Q}\, \bbox{D}\overline{U}\bbox{D}^T \bbox{\Delta g}_0 \, 
  [e_1(q)+e_2(q),G_{\rm BS}^{-1}]\, \bbox{\cal Q}
\nonumber\\  
&& - \bbox{\cal Q} \bbox{D}\overline{U}\bbox{D}^T [e_1(q)+e_2(q),
{\textstyle\frac{1}{2}}\bbox{\Delta g}_0 ]
   \bbox{D}\overline{U}\bbox{D}^T \bbox{\cal Q}\, , \\
q_\mu \bbox{J}^\mu_{\rm ex,eff} &=&
  \bbox{\cal Q}[e_1(q)+e_2(q),\bbox{D}\overline{U}\bbox{D}^T ]
\bbox{\cal Q} +
\bbox{\cal Q}\bbox{D}\overline{U}\bbox{D}^T
[e_1(q)+e_2(q),{\textstyle\frac{1}{2}}
    \bbox{\Delta g}_0 ] \bbox{D}\overline{U}\bbox{D}^T
\bbox{\cal Q} \, .
\end{eqnarray}
The terms quadratic in $\bbox{U}$ cancel when the two
equations are added, giving
\begin{eqnarray}
q_\mu \bbox{J}^\mu &=& \bbox{\cal Q}\,
[e_1(q)+e_2(q),2 G_{\rm BS}^{-1} + \bbox{D}\overline{U}\bbox{D}^T]
\bbox{\cal Q} 
  - \bbox{\cal Q}\, [e_1(q)+e_2(q),G_{\rm BS}^{-1}]\, 
 \bbox{\Delta g}_0 \bbox{D}\overline{U}\bbox{D}^T \bbox{\cal Q} 
 \nonumber\\ &&
 - \bbox{\cal Q}\, \bbox{D}\overline{U}\bbox{D}^T \bbox{\Delta g}_0 \, 
  [e_1(q)+e_2(q),G_{\rm BS}^{-1}]\, \bbox{\cal Q} \, . 
\label{divereff}
\end{eqnarray}
\narrowtext
This equation can be simplified using rules
(\ref{rule1}--\ref{rule5})
\begin{eqnarray}
q_\mu \bbox{J}^\mu
&=& \bbox{Q} [\bbox{e}(q),2 {\bbox{g}_0}^{-1} +
\bbox{D}\overline{U}\bbox{D}^T ]
\bbox{Q} \nonumber\\
&=&\left[ \bbox{e}(q),2{\bbox{g}_0}^{-1}\bbox{Q} +
\overline{\bbox{U}}\right]=
\left[ \bbox{e}(q),\bbox{g}^{-1}\right] \, ,
\label{WTspec2}
\end{eqnarray}
where we have introduced the matrix charge operator,
\begin{equation}
\bbox{e}(q)=\left(
\begin{array}{cc}
e_2(q) & 0 \\
0 & e_1(q)
\end{array}
\right)\, .
\end{equation}
Eq.~(\ref{WTspec2}) is the symmetric generalization of
Eq.~(\ref{WTspec}), and leads, together with the wave
Eq.~(\ref{symmwavee}), to the formal proof of gauge invariance.

\section{Current operators for the Gross equation}

In the previous section we derived Eq.~(\ref{GrossCurrent}) of the
current operator $J^\mu_{11}$, which is to be used for the
treatment of nonidentical particles where particle 1 is
on-shell, and the current operator $\bbox{J}^\mu$,
Eq.~(\ref{symmcur}), for use with identical particles.  In this
section we will show that, using the symmetry of the states,
(\ref{symmcur}) can be reduced to (\ref{GrossCurrent}) [with the
obvious requirement that the masses and charges are equal], so that
the form (\ref{GrossCurrent}) can be used in both cases.  We will
then decompose (\ref{GrossCurrent}) into individual terms and
give a diagrammatic interpretation of the current.  Finally, we
compare our results with the Bethe-Salpeter equation.

\subsection{Equivalence of the currents}

First, we recall the simplifications of the symmetric one-body
current terms given in Eqs.~(\ref{symm1}--\ref{symm4}).  Using
these results, the one body terms can be written in the following
form
\begin{figure}
\centerline{\epsfxsize=5in\epsffile{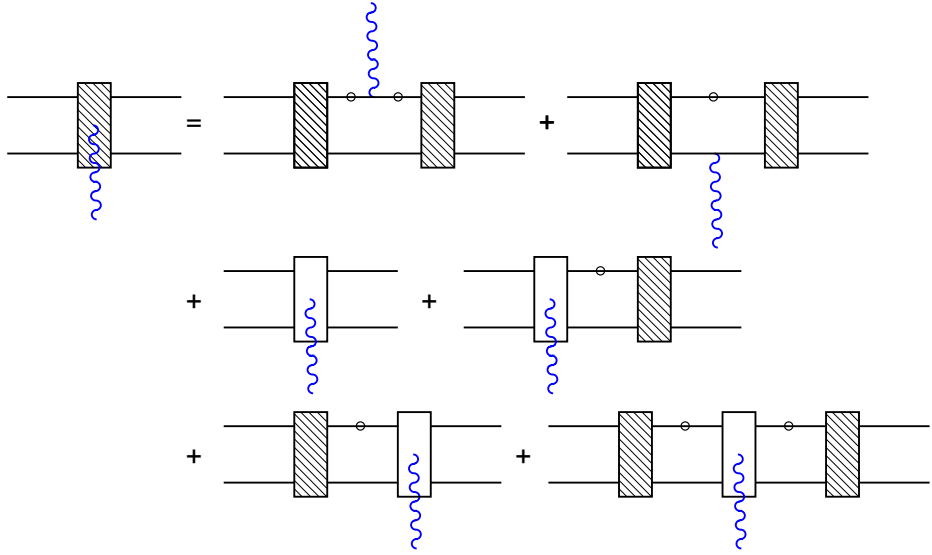}}
\caption{Feynman diagrams representing $\widetilde{J}^\mu_{\rm
int}$.  The open circles on particle line 1 are the difference
propagator $\Delta G_1$, the shaded rectangles are the quasipotential
$U$, and the open rectangles with photon attached are
$\overline{J}^\mu_{ex}$.}
\label{j_tilde}
\end{figure}
\widetext
\begin{equation}
\bbox{J}^\mu_{\rm IA,eff}  =\bbox{Q} \left(
\begin{array}{cc}
2 J^\mu_2 + j^\mu_1+j_1^{\mu\dagger} +j^\mu &
j^\mu_1+j_2^{\mu\dagger} +j^\mu \\ j^\mu_2+j_1^{\mu\dagger}
+j^\mu & 2 J^\mu_1 + j^\mu_2+j_2^{\mu\dagger} +j^\mu
\end{array}
\right)\bbox{Q}\, ,
\end{equation}
where
\begin{eqnarray}
&&j^\mu_1 = - J^\mu_1\Delta G_1 \overline{U}\nonumber\\
&&j_1^{\mu\dagger}= - \overline{U}\Delta G_1J^\mu_1\nonumber\\
&&j^\mu=-{\textstyle\frac{1}{2}}\,i \overline{U}
\Bigl[\Delta G_1 J_1^\mu
\Delta G_1 G_2 + G_2J^\mu_2G_2\Delta G_1 + (1\leftrightarrow2)
\Bigr]\overline{U}\, , \nonumber
\end{eqnarray}
\narrowtext
and $j_2$ is obtained from $j_1$ by substituting 2 for 1.
Next, recall that $\zeta{\cal P}_{12}$ exchanges particles 1 and 2
(where $\zeta=\pm$ depending on statistics of the particles), and
use the identities
\begin{eqnarray}
&&\left|\psi_2\right>=\zeta{\cal P}_{12}\left|\psi_1\right>
\nonumber\\
&&\Delta g_2=\zeta{\cal P}_{12}\,\Delta g_1\,\zeta{\cal P}_{12}
\nonumber\\
&&\overline{U}=\zeta{\cal
P}_{12}\,\overline{U}=\overline{U}\,\zeta{\cal P}_{12}=\zeta{\cal
P}_{12}\,\overline{U}\,\zeta{\cal P}_{12}
\nonumber\\
&&J^\mu_{\rm IA}=\zeta{\cal P}_{12}\,J^\mu_{\rm IA}\,\zeta{\cal
P}_{12}
\nonumber\\
&&\overline{J}^\mu_{\rm ex} =\zeta{\cal P}_{12}\,\overline{J}^\mu_{\rm ex} 
=\overline{J}^\mu_{\rm ex} \,\zeta{\cal P}_{12}
=\zeta{\cal P}_{12}\,\overline{J}^\mu_{\rm ex} \,\zeta{\cal P}_{12} 
\nonumber
\end{eqnarray}
to show that
\widetext
\begin{eqnarray}
&&\left< {\psi}_2\right|{J}^\mu_{1}
\left|{\psi}_2\right> = \left< {\psi}_1\right|{J}^\mu_{2}
\left|{\psi}_1\right>\nonumber\\
&&\left< {\psi}_1\right|{j}^\mu_{1}
\left|{\psi}_1\right>= \left< {\psi}_1\right|{j}^\mu_{1}
\left|{\psi}_2\right>= \left< {\psi}_2\right|{j}^\mu_{2}
\left|{\psi}_1\right>= \left< {\psi}_2\right|{j}^\mu_{2}
\left|{\psi}_2\right>\nonumber\\
&&\left< {\psi}_1\right|{j}^{\mu\dagger}_{1}
\left|{\psi}_1\right>= \left< {\psi}_2\right|{j}^{\mu\dagger}_{1}
\left|{\psi}_1\right>= \left< {\psi}_1\right|{j}^{\mu\dagger}_{2}
\left|{\psi}_2\right>= \left< {\psi}_2\right|{j}^{\mu\dagger}_{2}
\left|{\psi}_2\right>\nonumber\\
&&\left< {\psi}_1\right|{j}^\mu
\left|{\psi}_1\right>= \left< {\psi}_1\right|{j}^\mu
\left|{\psi}_2\right>= \left< {\psi}_2\right|{j}^\mu
\left|{\psi}_1\right>= \left< {\psi}_2\right|{j}^\mu
\left|{\psi}_2\right>\, . \nonumber
\end{eqnarray}
Hence
\begin{eqnarray}
\left< \bbox{\psi}\right|\bbox{J}^\mu
\left|\bbox{\psi}\right>&=& {\textstyle\frac{1}{4}}\,\Bigl[
\left< {\psi}_1\right|\bbox{J}^\mu_{11}\left|{\psi}_1\right>
+\left< {\psi}_1\right|\bbox{J}^\mu_{12}\left|{\psi}_2\right>
+\left< {\psi}_2\right|\bbox{J}^\mu_{21}\left|{\psi}_1\right>
+ \left<{\psi}_2\right|\bbox{J}^\mu_{22}\left|{\psi}_2\right>
\Bigr]\nonumber\\
&=&\left< \psi_1\right|\biggl[ J^\mu_2-J^\mu_1G_1\overline{U}
-\overline{U}G_1J^\mu_1
-i\overline{U}\Delta G_1J^\mu_1\Delta G_1G_2\overline{U}
\nonumber \\
& & \qquad -i\overline{U}\Delta G_1G_2J^\mu_2G_2\overline{U}
+ \left( 1-\overline{U}\Delta g_1 \right)
\overline{J}^\mu_{\rm ex}
\left( 1-\Delta g_1\overline{U} \right)\biggr]\left| \psi_1\right> .
\label{cureffs}
\end{eqnarray}
\narrowtext
Note that this is identical to (\ref{GrossCurrentME}), provided
that the symmetrized two-body interaction is substituted for
$U$, the symmetrized interaction current is substituted for
$J^\mu_{\rm ex}$, and the masses and charges are set equal to
each other.  Hence we have show that Eq.~(\ref{GrossCurrentME})
may be used either for identical or nonidentical particles.
 
\subsection{Final expressions for the currents}

Now we will write explicit expressions for the
matrix elements of the effective current between two bound states
and between a bound initial state and a scattering final state. To
facilitate this define an effective {\it interaction\/} current
\widetext
\begin{equation}
\widetilde{J}^\mu_{\rm int}=
-i\overline{U}\Delta G_1J^\mu_1\Delta G_1G_2\overline{U}
 -i\overline{U}\Delta G_1G_2J^\mu_2G_2\overline{U}
+ \left( 1-\overline{U}\Delta g_1 \right)
\overline{J}^\mu_{ex}
\left( 1-\Delta g_1\overline{U} \right)\, .
\label{curtilde}
\end{equation}
\narrowtext
This current is illustrated diagrammatically in Fig.~\ref{j_tilde}.
\begin{figure}
\centerline{\epsfxsize=5in\epsffile{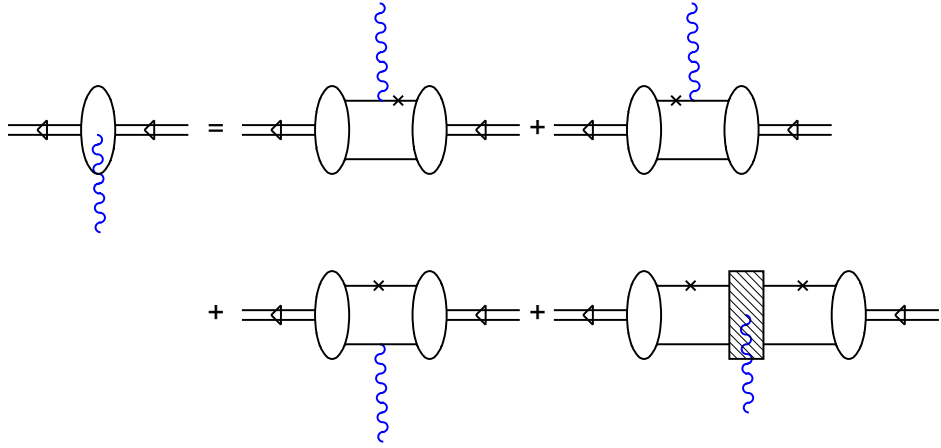}}
\caption{Feynman diagrams representing the matrix element of the effective
current between bound states.}\label{elastic}
\end{figure}

The matrix element of the effective current between bound states
can then be written
\widetext
\begin{eqnarray}
\left< \bbox{\psi}\right|\bbox{J}^\mu
\left|\bbox{\psi}\right>&=&
\left< \psi_1\right|\left[ J^\mu_2-J^\mu_1G_1\overline{U}
-\overline{U}G_1J^\mu_1
+\widetilde{J}^\mu_{\rm int}\right]\left| \psi_1\right>
\nonumber\\
&=&
\left< \psi_1\right| J^\mu_2\left| \psi_1\right>-
\left< \psi_1\right| J^\mu_1G_1\overline{U}g_1\left| \Gamma_1\right>
-\left< \Gamma_1\right| g_1\overline{U}G_1J^\mu_1\left| \psi_1
\right> +\left< \psi_1\right|\widetilde{J}^\mu_{\rm int}\left|
\psi_1\right> .
\label{elasticME}
\end{eqnarray}
In numerical calculations it is often convenient to introduce an
off-shell vertex function
\begin{equation}
\left| \Gamma \right>=-\overline{U}g_i\left| \Gamma_i\right> \, ,
\label{Gammaos}
\end{equation}
where $i= 1,2$.  This can be used to rewrite (\ref{elasticME}) as
\begin{equation}
\left< \bbox{\psi}\right|\bbox{J}^\mu
\left|\bbox{\psi}\right>=
\left< \Gamma\right| G_1 J^\mu_1\left| \psi_1\right>
+\left< \psi_1\right| J^\mu_1 G_1\left| \Gamma\right>
+\left< \psi_1\right| J^\mu_2\left| \psi_1\right>
+\left< \psi_1\right|\widetilde{J}^\mu_{\rm int}\left|
\psi_1\right> .
\label{elasticME2}
\end{equation}
\narrowtext
The Feynman diagrams representing the elastic matrix element are shown in
Fig.~\ref{elastic}.
%
\begin{figure}
\centerline{\epsfxsize=5in\epsffile{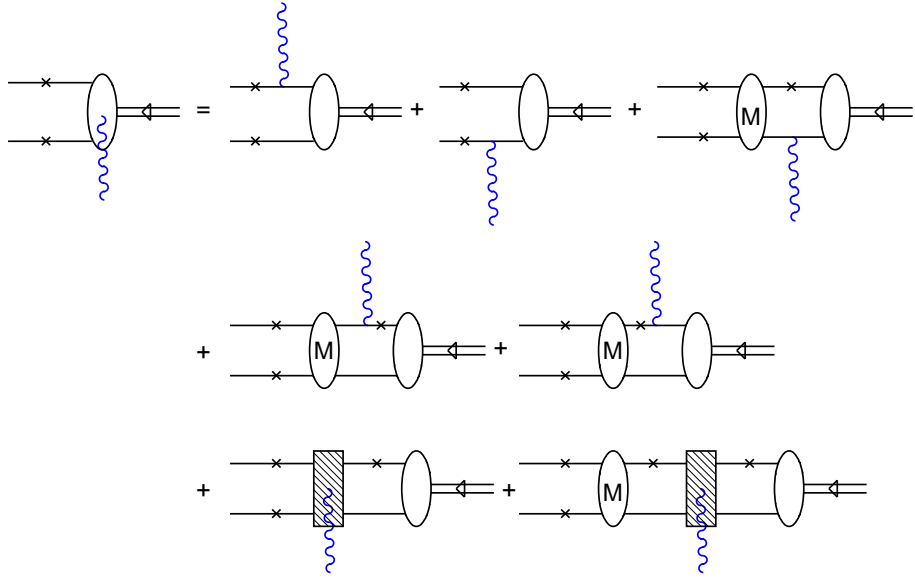}}
\caption{Feynman diagrams representing the matrix element of the effective
current between a final scattering state and an initial bound state.}
\label{inelastic}
\end{figure}

The matrix element of the effective current between a bound initial
state and a scattering final state is
\widetext
\begin{eqnarray}
\left< \bbox{\psi}^{(-)}\right|\bbox{J}^\mu
\left|\bbox{\psi}\right>&=&
\left< \bbox{p}_1,s_1;\bbox{p}_2,s_2 \right|{\cal A}_2  (1-Mg_1) {\cal Q}_1
\left[ J^\mu_2-J^\mu_1G_1\overline{U}
-\overline{U}G_1J^\mu_1
+\widetilde{J}^\mu_{\rm int}\right]\left| \psi_1\right>
\label{inelasticME}
\end{eqnarray}
Using the identities
\begin{eqnarray}
&&{\cal A}_2 M=  M \nonumber\\
&&{\cal A}_2 \widetilde{J}^\mu_{\rm int}
= \widetilde{J}^\mu_{\rm int} \nonumber\\
&&\left< \bbox{p}_1,s_1;\bbox{p}_2,s_2 \right| {\cal A}_2 
 \left(1-Mg_1 \right){\cal Q}_1 \overline{U} 
  = \left< \bbox{p}_1,s_1;\bbox{p}_2,s_2 \right| M 
\nonumber\\
&&\left< \bbox{p}_1,s_1;\bbox{p}_2,s_2 \right| {\cal A}_2  
\left(J^\mu_2-J^\mu_1 G_1\overline{U}\right) \left|\psi_1\right>
\nonumber\\
&& \quad \quad  
 =\left< \bbox{p}_1,s_1;\bbox{p}_2,s_2 \right| {\textstyle\frac{1}{2}}
 \Bigl[ J^\mu_2 G_2 \left|\Gamma_1\right> + J^\mu_1 G_1 \left|\Gamma_2\right>
 + J^\mu_2 G_2 \left|\Gamma\right> + J^\mu_1 G_1 \left|\Gamma\right> \Bigr] 
\nonumber\\ 
&& \quad \quad \quad
\left< \bbox{p}_1,s_1;\bbox{p}_2,s_2 \right| 
 \Bigl[ J^\mu_1 G_1 \left|\Gamma_2\right> + J^\mu_2 G_2 \left|\Gamma_1\right> 
\Bigr] = \left< \bbox{p}_1,s_1;\bbox{p}_2,s_2 \right|
\Bigl[ J^\mu_1 \left|\psi_2\right> + J^\mu_2  \left|\psi_1\right> \Bigr]
\nonumber
\end{eqnarray}
where to get the last relation we employed (\ref{Gammaos}), 
the current matrix element can be rewritten
\begin{eqnarray}
\left< \bbox{\psi}^{(-)}\bigl|\bbox{J}^\mu
\bigr|\bbox{\psi}\right>&=&
\Bigl<\bbox{p}_1,s_1;\bbox{p}_2,s_2 \Bigl| \biggl[
 J^\mu_1 \left|\psi_2\right>+ J^\mu_2 \left|\psi_1\right>
- M g_1 J^\mu_2  \left|\psi_1\right> - M G_1 J^\mu_1 \left| \psi_1\right>
\nonumber\\
&&\qquad\qquad\quad  
 + M g_1 Q_1 J^\mu_1 G_1 \overline{U}\left|\psi_1\right> 
+ (1-M g_1 Q_1 )\widetilde{J}^\mu_{\rm int} \left|\psi_1\right> \biggr]
\nonumber\\
&=&
\Bigl< \bbox{p}_1,s_1;\bbox{p}_2,s_2 \Bigl|\biggl[
J^\mu_1 \left|\psi_2\right>+ J^\mu_2 \left|\psi_1\right>
- M G_2 J^\mu_2 \left|\psi_1\right> - M G_1 J^\mu_1 \left|\psi_1\right>
\nonumber\\
&&\qquad\qquad\quad  - M  Q_1 J^\mu_1 G_1 G_2 \left|\Gamma\right>
+ (1-M Q_1 G_2 )\widetilde{J}^\mu_{\rm int}\left|\psi_1\right> \biggr]
\, . \label{inelasticME2}
\end{eqnarray}
\narrowtext 
The Feynman diagrams representing the inelastic matrix element are shown in
Fig.~\ref{inelastic}.


\subsection{Comparison to Bethe-Salpeter matrix elements}

 Let us now compare the matrix elements derived above with those
of the Bethe-Salpeter description. First, consider the elastic
Bethe-Salpeter matrix element
\widetext
\begin{eqnarray}
 \left< \psi \right| J^\mu \left| \psi \right>_{\rm BS} &=&
 \left< \Gamma \right| G_{\rm BS} \left( i J^\mu_1 G^{-1}_2+
  iJ^\mu_2G^{-1}_1+\overline{J}^\mu_{\rm ex}
  \right)  G_{\rm BS} \left| \Gamma \right> \nonumber\\
  &=& \left< \Gamma \right| -i G_1 J^\mu_1 G_1 G_2
 -i G_1 G_2 J^\mu_2 G_2 -
 G_1 G_2 \overline{J}^\mu_{\rm ex}  G_1 G_2 \left| \Gamma \right>
\, . \label{elasticBS}
\end{eqnarray}
With the help of simple identities obtained from
Eqs.~(\ref{sepofprop}) and (\ref{rule6})
\begin{eqnarray}
  -i G_1 J^\mu_1 G_1 G_2 &=& {\cal Q}_1 J^\mu_1 G_1 G_2 + G_1 J^\mu_1
{\cal Q}_1 G_2
                   -i \Delta G_1 J^\mu_1 \Delta G_1 G_2 \, ,
\nonumber\\
  -i G_1 G_2 J^\mu_2 G_2  &=& {\cal Q}_1 G_2 J^\mu_2 G_2
              -i \Delta G_1 G_2 J^\mu_2 G_2 \, , \nonumber\\
  - G_1 G_2 \overline{J}^\mu_{\rm ex}  G_1 G_2    &=&
 ({\cal Q}_1 - i \Delta G_1) G_2 \overline{J}^\mu_{\rm ex}
 G_2 ({\cal Q}_1 - i \Delta G_1)  \, ,
\label{curriden}
\end{eqnarray}
relation (\ref{Gammaos}), and the conventions (\ref{Qconvention}) and
(\ref{Qrelation}), we can rewrite (\ref{elasticBS}) as
\begin{eqnarray}
 \left< \psi \right| J^\mu \left| \psi \right>_{\rm BS}  =&&
\left< \Gamma \right| Q_1 G_2
  \left[ J^\mu_2- J^\mu_1 G_1\overline{U}
-\overline{U}G_1J^\mu_1
+\widetilde{J}^\mu_{\rm int} \right]  Q_1 G_2 \left| \Gamma
\right> \nonumber\\
=&& \left< \psi_1 \right|
  \left[ J^\mu_2- J^\mu_1 G_1\overline{U}
-\overline{U}G_1J^\mu_1
+\widetilde{J}^\mu_{\rm int} \right]  \left| \psi_1\right>
 \, , \label{elasticBScom}
\end{eqnarray}
\narrowtext
with $\widetilde{J}^\mu_{\rm int}$ defined by (\ref{curtilde}).
We have demonstrated again that our spectator matrix element
(\ref{elasticME}) exactly equals Bethe-Salpeter one
(\ref{elasticBS}).
Of course, this should be so  by construction.
Still, the derivation of this section
gives a useful shortcut to the correct spectator matrix element.
It  also illustrates how the current $J^\mu_2-
J^\mu_1G_1\overline{U} -\overline{U}G_1J^\mu_1$ in
(\ref{elasticME}) and (\ref{elasticBScom}) follows from the
Bethe-Salpeter matrix element if one puts the first particle
on-shell, i.e., if one keeps only the first term in
Eq.~(\ref{matrixquasi}) for the  quasipotential (for
consistency we should replace $\overline{U} \to \overline{V}$,
which also holds when all terms with $\Delta G_1$ are omitted).
The last part of the effective current $\widetilde{J}^\mu_{\rm int}
$ then gathers all higher order effects. A very similar
consideration can be applied to the break-up matrix element
(\ref{inelasticME}).  One finds that the parts of the matrix
element with loops can again be related by identities
(\ref{curriden}), while the  loopless parts, i.e., an IA
contributions without a final-state interaction, are identical for
both approaches.


\section{Charge conservation}

In this section we show that the total charge of a bound state is
equal to the sum of the charges of its constituents, $e_1 +
e_2$, and discuss how this result emerges automatically in the
Bethe-Salpeter and spectator formalisms. In this
discussion we will assume for definiteness that the two particles
are nonidentical, but our results will hold for identical
particles also since the current operator in the latter case is obtained
by  symmetrization of the former one.

First recall that taking the $q \to 0$ limit of the one-body
Ward-Takahashi (WT) identity, Eq.~(\ref{Ward1}), implies that the
one-body currents satisfy
\begin{equation}
 J^\mu_i (0) = - e_i
 \frac{d G_i^{-1} (p_i)}{d p_{i,\mu }} \, .
\label{cur1_0}
\end{equation}
This relation will be used in both formalisms.

Next consider the two-body WT identity, Eq.~(\ref{Ward2}), in the
context of the BS formalism.  It is well known \cite{r1} that the
contribution to the charge operator which comes from the exchange
current can be uniquely determined by taking the $q \to 0$ limit of
(\ref{Ward2}). The derivation of this result was discussed in
great detail by  Bentz \cite{b1}, and we only review it briefly
here.  Since the overall four-momentum is conserved, the kernel is
a function of only three independent four-momenta, which can be
chosen to be either $P$, $p$, and $p'$, or $P$, $p_1$, and $p'_1$.
Depending on how we choose the independent momenta, the $q \to 0$
limit of Eq.~(\ref{Ward2}) gives
\widetext
\begin{eqnarray}
 J^\mu_{\rm ex} (0) &=& - (e_1+ e_2) \, \frac{\partial
 V(p',p,P)}{\partial P_{\mu }} - \frac{(e_1-e_2)}{2} \,
 \left[\frac{\partial  V(p',p,P)}{\partial p'_{\mu }}
+ \frac{\partial V(p',p,P)}{\partial p_{\mu}}\right]
\label{curex0a} \\
&=& - (e_1+ e_2) \, \frac{\partial
 V(p'_1,p_1,P)}{\partial P_{\mu }} - e_1\,
 \left[\frac{\partial  V(p_1',p_1,P)}{\partial p'_{1,\mu }}
+ \frac{\partial V(p'_1,p_1,P)}{\partial p_{1,\mu}}\right]
\, ,
\label{curex0}
\end{eqnarray}
where in (\ref{curex0a}) the partial derivative with respect $P_\mu$
implies that the independent vectors $p_\mu$ and $p'_\mu$ are held
constant, while in (\ref{curex0}) the partial derivative with respect
$P_\mu$ implies that the independent vectors $p_{1,\mu}$ and
$p'_{1,\mu}$ are held constant. Similarly,
\begin{eqnarray}
-i\frac{d G_1^{-1} (p_1)}{d p_{1,\mu }}G_2^{-1}(p_2) =&& -
\frac{\partial G_{\rm BS}^{-1} (p,P)}{\partial P_{\mu }} - {1\over2}
\frac{\partial G_{\rm BS}^{-1} (p,P)}{\partial p_{\mu }} \label{propxx1}\\
=&& - \frac{\partial G_{\rm BS}^{-1} (p_1,P)}{\partial P_{\mu }} -
\frac{\partial G_{\rm BS}^{-1} (p_1,P)}{\partial p_{1,\mu }}
\label{propxx2}
\end{eqnarray}
and
\begin{eqnarray}
-iG_1^{-1}(p_1)\frac{d G_2^{-1} (p_2)}{d p_{2,\mu }} =&& -
\frac{\partial G_{\rm BS}^{-1} (p,P)}{\partial P_{\mu }} + {1\over2}
\frac{\partial G_{\rm BS}^{-1} (p,P)}{\partial p_{\mu }} \label{propxx3}\\
=&&- \frac{\partial G_{\rm BS}^{-1} (p_1,P)}{\partial P_{\mu }}
\label{propxx4} \, .
\end{eqnarray}
\narrowtext
The correct forms of these equations depend on our
choice of independent vectors.  In the BS case either of the forms can be
used since there are no additional constraints on the vectors, but in
the spectator case with particle 1 on-shell we must use
(\ref{curex0}), (\ref{propxx2}), and (\ref{propxx4})  because  $p_2$
will explicitly depend on $P$ in cases when $p_1$ is constrained.

We use (\ref{curex0a}),  (\ref{propxx2}), and
(\ref{propxx4}) to evaluate the bound state matrix element of the
charge operator in the BS formalism
\widetext
\begin{eqnarray}
\left< \psi \right| J^\mu (0) \left| \psi \right>  &=&-
\left< \psi \right|
  e_1\, i \frac{d G_1^{-1} (p_1)}{d p_{1,\mu }}G_2^{-1}(p_2)
 + e_2\, i G_1^{-1}(p_1) \frac{d G_2^{-1} (p_2)}{d p_{2,\mu }}
 - J^\mu_{\rm ex} (0) \left| \psi \right> \nonumber\\
&=& -\left< \psi \right|  (e_1+ e_2)\, \biggl(
 \frac{\partial G_{\rm BS}^{-1}(p,P)}{\partial P_{\mu }} +
  \frac{\partial V(p',p,P)}{\partial P_{\mu }} \biggr)   \nonumber\\
&& \quad  + \frac{e_1-e_2}{2} \biggl(
 \frac{\partial G_{\rm BS}^{-1}(p,P)}{\partial p_{\mu }}
 + \frac{\partial V(p',p,P)}{\partial p'_{\mu }}
 + \frac{\partial V(p',p,P)}{\partial p_{\mu }}
 \biggr)  \left| \psi \right>   \nonumber\\
 &=& (e_1+e_2)\, 2 P^\mu \, ,
\label{bsj0}
\end{eqnarray}
\narrowtext
where the normalization condition for the BS vertex function
\cite{norm,r1} was used in the last step to simplify the
$(e_1+ e_2)$ terms, and the cancellation of the $(e_1- e_2)$ terms
follows from integrating $\partial G^{-1}_{\rm BS}/\partial p_\mu$
by parts and using the  bound state BS equation (\ref{wave_eqn_bs}). The
final form of (\ref{bsj0}) shows that the charge is conserved.

Now we turn to the spectator formalism with the effective current
given in Eq.~(\ref{GrossCurrent}).
We begin by pointing out that, unlike in the
Bethe-Salpeter case, one  cannot  obtain $J^\mu_{11}(0)$
from the corresponding Ward-Takahashi identity (\ref{WTspec}).  For
nonidentical particles the clear indication of this fact is that
the charge of the first particle (which can be completely
arbitrary) is absent from the WT relation (\ref{WTspec}) and
any current determined from this relation would therefore depend
on $e_2$ only, which is certainly not correct.  The reason for
this was alluded to in Sec.~III:  the condition
restricting the first particle to its mass shell leads to an
effective current in which the terms proportional to the charge of
particle 1 are purely {\it transverse\/}. There can be also
transverse currents in the BS case, but they are of the form
$a_{\mu \nu} q^\nu$, with $a_{\mu \nu}$ antisymmetric and
nonsingular for $q \to 0$,
and hence they vanish in this limit, and all parts of the current
contributing to the charge can be recovered from the WT identity
(see Ref.~\cite{b1}). In the spectator formalism those parts of
the current which are transverse by virtue of the on-shell condition
do {\it not vanish\/} in the $q \to 0$ limit. Therefore, the effective
current in $q \to 0$ limit cannot be fully recovered from the WT
identity and has to be obtained
by taking the limit  explicitly.

The effective spectator current for zero photon momentum
follows from Eqs.~(\ref{GrossCurrent}) and (\ref{curex0})
\widetext
\begin{eqnarray}
 J^\mu_{11}(0) &=& {\cal Q}_1 \biggl\{ - e_2 \frac{d G_2^{-1}}{d
 p_{2,\mu }} - \left[J_1^\mu G_1 U + U G_1 J_1^\mu
\right]_{q \,\to\, 0}\,
+ \left( 1- U \Delta g_1 \right) J^\mu_{\rm ex} (0)
\left( 1- \Delta g_1  U \right) \nonumber\\
&& +\left[ ie_1\, U \Delta G_1
 \frac{d G_1^{-1}}{d p''_{1,\mu }}
\, \Delta G_1 G_2 U     + ie_2\, U \Delta G_1 G_2
 \frac{d G_2^{-1}}{d p''_{2,\mu }}
\, G_2 U  \right] \, \biggr\}  {\cal Q}_1  \, .
\label{longeq}
\end{eqnarray}
The first term in the last line of this equation can be reduced if we
use rules (\ref{rule3}) and (\ref{rule4}) and integrate by parts twice
(noting that $p_1$ is unconstrained in this loop and that $P$ is to be
held constant)
\begin{eqnarray}
i e_1 U \Delta G_1
 \frac{d G_1^{-1}}{d p''_{1,\mu }} \, \Delta G_1 G_2 U =&&
-i e_1 \frac{\partial {U}(p_1',p_1'',P)}{\partial p''_{1,\mu}}
\Delta G_1 G_2 U - i e_1 U
\Delta G_1 G_2 \frac{\partial {U}(p_1'',p_1,P)}{\partial
p''_{1,\mu}}\nonumber\\
&&  -2ie_1 U \frac{\partial \Delta G_1}{\partial p''_{1,\mu}}
G_2 U -ie_1 U \Delta G_1 \frac{\partial G_2}{\partial p''_{1,\mu}}  U
\nonumber\\
=&& e_1 U \frac{\partial \Delta g_1}{\partial p''_{1,\mu}}U + ie_1 U
\Delta G_1 \frac{\partial G_2}{\partial p''_{1,\mu}}  U \, ,
\end{eqnarray}
\narrowtext
where here and below $p'_1$ and $p_1$ are the four-momenta of particle 1
after and before the interaction, respectively, and $p''_1$ denotes the
momenta of the loop integration implied by the product $U \dots  U$.
Using
\begin{equation}
 G_2\, \frac{d G_2^{-1}(p_2)}{d p_{2,\mu }}\, G_2
 = \frac{\partial G_2(P-p_1)}{\partial p_{1,\mu }} \, , \label{proprel}
\end{equation}
the second term in the last line of Eq.~(\ref{longeq}) becomes
\begin{equation}
 i e_2 U \Delta G_1 G_2
 \frac{d G_2^{-1}}{d p''_{2,\mu }}
\, G_2 U = i e_2 U \Delta G_1 \frac{\partial G_2}{\partial
p''_{1,\mu }} U \,  .
\end{equation}
The term with the exchange current $J^\mu_{\rm ex} (0)$ is simplified
with the help of
\begin{eqnarray}
 \frac{\partial U}{\partial p'_{1,\mu}} &=&
 \frac{\partial V}{\partial p'_{1, \mu}} (1 - \Delta g_1 \, U)
\, , \label{udera}\\
\frac{\partial U}{\partial p_{1, \mu}} &=&  (1 - U\, \Delta g_1 )
\frac{\partial V}{\partial p_{1, \mu}} \, .
\label{uder}
\end{eqnarray}
 These relations are obtained
by differentiating the corresponding off-shell quasipotential equations
and using the fact that the structure of the integral equations insures
that the only dependence on the final momentum $p'_1$ in
(\ref{udera}), or  on the initial momentum $p_1$ in (\ref{uder}),
is found in the kernel $V$. A similar argument gives
\begin{equation}
 \frac{\partial U}{\partial P_{\mu}} =
 \frac{\partial V}{\partial P_{\mu}} (1 - \Delta g_1 \, U)
-V \frac{\partial (\Delta g_1 U)}{\partial P_{\mu}}
\end{equation}
and hence
\widetext
\begin{eqnarray}
\left( 1- U \Delta g_1 \right)  \frac{\partial
{V}}{\partial P_{\mu }} \left( 1- \Delta g_1  U \right) =&&
 \frac{\partial U}{\partial P_{\mu}} -
 U\Delta g_1 \frac{\partial U}{\partial P_{\mu}} +
 U \frac{\partial (\Delta g_1 U)}{\partial P_{\mu}}
\nonumber\\
=&& \frac{\partial U}{\partial P_{\mu}} + U \frac{\partial \Delta
g_1}{\partial P_{\mu}} U \, .\label{tripleu}
\end{eqnarray}
Hence using (\ref{curex0}), (\ref{udera}), (\ref{uder}), and
(\ref{tripleu}) gives
\begin{eqnarray}
\left( 1- U \Delta g_1 \right) &J^\mu_{\rm ex}& (0)
\left( 1- \Delta g_1  U \right) \nonumber\\
&=& -(e_1+ e_2) \,\left[ \frac{\partial U}{\partial P_{\mu}} + U
\frac{\partial \Delta g_1}{\partial P_{\mu}} U \right]
-e_1 \,\left[\frac{\partial {U}(p_1',p_1,P)} {\partial
p'_{1,\mu }} +
\frac{\partial {U}(p_1',p_1,P)}{\partial p_{1,\mu}}\right]
\nonumber\\
&&+ e_1 \,\left[U \Delta g_1
\frac{\partial {U}(p_1'',p_1,P)}{\partial p''_{1,\mu }} +
\frac{\partial {U}(p_1',p_1'',P)}{\partial p''_{1,\mu}}
\Delta g_1 U   \right]  \label{jex0a}\\
&=& -(e_1+ e_2) \,\left[ \frac{\partial U}{\partial P_{\mu}} + U
\frac{\partial \Delta g_1}{\partial P_{\mu}} U \right]
-e_1 \,\left[\frac{\partial {U}} {\partial
p'_{1,\mu }} +
\frac{\partial {U}}{\partial p_{1,\mu}}
+ U \frac{\partial \Delta g_1}{\partial p''_{1,\mu}} U \right] \, ,
\label{jex0}
\end{eqnarray}
\narrowtext
Since $p''_1$ is off-shell in the integration loop,
we could integrate by parts to simplify the last line
of (\ref{jex0a}).

Making these substitutions and combining terms permits us to simplify
Eq.~(\ref{longeq})
\widetext
\begin{eqnarray}
 J^\mu_{11}(0) = Q_1 \biggl\{ && - e_2 \frac{\partial G_2^{-1}}{\partial
 P_{\mu }} -\left[J_1^\mu G_1 U + U G_1 J_1^\mu
\right]_{q \,\to\, 0} -(e_1+ e_2) \,\frac{\partial U} {\partial
P_{\mu}} \nonumber\\
&& -e_1 \,\left(\frac{\partial {U}(p_1',p_1,P)} {\partial
p'_{1,\mu }} + \frac{\partial {U}(p_1',p_1,P)}{\partial p_{1,\mu}}
\right)\nonumber\\
&& +(e_1+ e_2) \,U\,\left[i \Delta G_1 \frac{\partial
G_2}{\partial p''_{1,\mu }}  - \frac{\partial \Delta g_1}{\partial
P_{\mu}}  \right] \,U\, \biggr\}  Q_1  \, .
\label{longeq2}
\end{eqnarray}
Recall that partial derivative with respect to $P$ holds $p_1$
constant, and hence
\begin{equation}
{\partial \Delta g_1(p_1,P)\over \partial P_\mu}= -i\Delta G_1(p_1)
{\partial G_2(P-p_1)\over \partial P_\mu}= i\Delta G_1(p_1)
{\partial G_2 (P-p_1)\over \partial p_{1,\mu}}\, ,
\end{equation}
\narrowtext
so the last line proportional to $U^2$ in (\ref{longeq2})
cancels. The remaining terms will be now shown to be
proportional to the normalization condition. Let us point out
that exactly such terms (with $U \to V$) would appear if only
the leading order quasipotential with corresponding currents are
considered.

To simplify (\ref{longeq2}) one has to reduce
the term $\left[J_1^\mu G_1 U + U G_1 J_1^\mu \right]_{q \,\to\, 0}$.
It must be treated
carefully as each term is singular as $q\,\to\,0$, but, as discussed
in Ref~\cite{norm}, the singularities cancel in the sum.  Following
the argument developed in Ref.~\cite{norm}, using the notation
$\hat{p}_1$ and $\hat{p}'_1$ to indicate those cases where the
four-momenta of particle 1 are restricted to their mass shell,
and exploiting  the bound state equation (\ref{boundseq})
gives
\widetext
\begin{eqnarray}
- \left< \Gamma_1 \right| &&G_2
 \bigl[J_1^\mu  G_1 U + U G_1 J_1^\mu \bigr]_{q\,\to\,0}\, G_2
\left| \Gamma_1 \right>  \nonumber\\
&&= \left< \Gamma_1 \right| G_2 \left[ U_{11} G_2 J_1^\mu G_1 U +
  U G_1 J_1^\mu G_2 U_{11}\right]  G_2 \left|\Gamma_1
\right>_{q\,\to\,0}
\nonumber\\
&&\simeq - {e_1\over q_\mu} \left< \Gamma_1 \right| G_2 \biggl\{
  U_{11}(\hat{p}'_1,\hat{p}''_1,P+q)
 \left[ G_1^{-1}(\hat{p}''_1)- G_1^{-1}(\hat{p}''_1-q)\right]
 G_1(\hat{p}''_1-q) \nonumber\\
&&\qquad \qquad \times G_2(P-\hat{p}''_1+q) U (\hat{p}''_1-q,\hat{p}_1,P)
   + U(\hat{p}'_1,\hat{p}''_1+q,P+q) G_2 (P-\hat{p}''_1) \nonumber\\
&&\qquad \qquad  \times G_1(\hat{p}''_1+q)
 \left[ G_1^{-1}(\hat{p}''_1+q) - G_1^{-1}(\hat{p}''_1) \right]
 U_{11}(\hat{p}''_1, \hat{p}_1,P) \biggr\}  G_2
 \left| \Gamma_1 \right>_{q\,\to\,0} \nonumber\\
&&={e_1\over q_\mu} \left< \Gamma_1 \right| G_2 \biggl\{
  U_{11}(\hat{p}'_1,\hat{p}''_1,P+q) G_2(P-\hat{p}''_1+q)
  U (\hat{p}''_1-q,\hat{p}_1,P)  \nonumber\\
&&\qquad \qquad - U(\hat{p}'_1,{p}''_1+q,P+q)
  G_2 (P-\hat{p}''_1) U_{11}(\hat{p}''_1, \hat{p}_1,P)
\biggr\} G_2 \left| \Gamma_1 \right>_{q\,\to\,0} \nonumber\\
&&= e_1\, \left< \Gamma_1 \right| G_2 \biggl(
  U_{11} {\partial G_2 \over \partial P_{\mu}} U_{11}
  - {\partial U  \over \partial \hat{p}''_{1,\mu}} G_2 U_{11} -
  U_{11}  G_2 {\partial U  \over \partial \hat{p}''_{1,\mu}}  \biggr) G_2
  \left| \Gamma_1 \right> \nonumber\\
&& =  e_1\, \left< \Gamma_1 \right|
 {\partial G_2 (P-\hat{p}_1 ) \over \partial P_{\mu}} \nonumber\\
&& \qquad \qquad + G_2(P-\hat{p}'_1)\, \left(
 \frac{\partial U(\hat{p}'_1,\hat{p}_1,P)}{\partial \hat{p}'_{1,\mu }} +
 \frac{\partial U(\hat{p}'_1,\hat{p}_1,P)}{\partial \hat{p}_{1,\mu }} \right)
 \,  G_2 (P-\hat{p}_1)\,  \left| \Gamma_1 \right>   \, ,
\label{limit1}
\end{eqnarray}
\narrowtext
where, in going from the second to the third step we used
$G_1^{-1}Q_1=0$ for the on-shell momentum $\hat{p}''_1$, and in
the last step we used the bound state equation to remove the
factors of $G_2U$ whenever possible. When simplifying (\ref{limit1})
it is important to choose the dummy integration momentum
$p''_1$ so that the on-shell condition does not depend on the photon
momentum $q$. Finally, substituting this result
into (\ref{longeq2})
gives
\begin{equation}
 J^\mu_{11}(0) = (e_1+e_2)\, Q_1 \left\{ G_2^{-1} \frac{\partial
G_2}{\partial P_{\mu }} G_2^{-1} - \frac{\partial U} {\partial P_{\mu}}
\right\} Q_1 \, ,
\label{final3}
\end{equation}
and the elastic matrix element of the effective spectator
current at $q=0$ becomes
\widetext
\begin{eqnarray}
\left< \psi_1 \right|  J^\mu_{11}(0)  \left| \psi_1 \right>=&&
\left< \Gamma_1 \right| G_2  J^\mu_{11}(0) G_2
\left| \Gamma_1 \right> \nonumber\\
=&& (e_1+e_2) \left< \Gamma_1 \right|  \left(
 \frac{\partial G_2(P-\hat{p}_1)}{\partial P_{\mu }} \right. \nonumber\\
&& \left. - G_2 (P-\hat{p}'_1)\, \frac{\partial U(\hat{p}'_1,\hat{p}_1,P)}
 {\partial P_{\mu }}\,  G_2 (P-\hat{p}_1)\,  \right) \left|
\Gamma_1 \right>   \, .
\label{curreffel}
\end{eqnarray}
However,  the normalization condition for the spectator vertex
function is just
\begin{equation}
 \left< \Gamma_1 \right|  \left(
 \frac{\partial G_2(P-\hat{p}_1)}{\partial P_{\mu }}
 - G_2 (P-\hat{p}'_1)\, \frac{\partial U(\hat{p}'_1,\hat{p}_1,P)}
 {\partial P_{\mu }}\,  G_2 (P-\hat{p}_1)\,  \right)
\left| \Gamma_1 \right> = 2 P^\mu \, .
\label{gammanor}
\end{equation}
\narrowtext
This was discussed in great detail in \cite{norm}, where it was derived
from the nonlinear form of the spectator equation without reference to
the e.m.~current (in that reference the spectator kernel was denoted by
$V$, but the derivation did not specify the kernel in any way and holds
equally well for the kernel $U$).
Obviously the relations (\ref{curreffel}) and (\ref{gammanor}) are
consistent with
\begin{equation}
 \left< \Gamma_1 \right| G_2  J^\mu_{11}(0) G_2 \left| \Gamma_1 \right>
  = (e_1+e_2) \, 2 P^\mu \, ,
\end{equation}
which is the statement that the charge of the bound state is
$e_1+e_2$, completing our proof.

Our derivation is valid for any interaction $V$ and the corresponding
quasipotential $U$  (e.g., also for phenomenological ones, such as a
separable interaction).  It is only  necessary to have an interaction
current at the Bethe-Salpeter level consistent with the one body
current, so that the total BS current is conserved.  Furthermore,
since we have not specified the spins of the constituents or the bound
state in  our  derivation, it should apply for arbitrary spins.


\section{Truncation}

To this point, no approximations have been made in constructing either
the n-point functions or the effective current operators. In particular,
the equivalence between the Gross equation and its quasipotential
and the Bethe-Salpeter equation is exact only if the Bethe-Salpeter
kernel $V$ and the spectator kernel $U$ are related by
Eq.~(\ref{quasipotential}).  This means that if one of the kernels is
truncated to some finite order, the other must involve terms of {\it
all\/} orders.  In practice, {\it both\/} kernels are generally truncated
to some finite order and the two formalisms do not give identical
results.  The usual approximation is to keep only the
one-boson-exchange-contribution, either for $V^{(1)}$ or $U^{(1)}$. The
problem is then to verify that the various relations leading to conserved
current matrix elements are maintained in the presence of the truncation.

First assume that we have some Bethe-Salpeter kernel
${V}\rightarrow\lambda{V}$ and the associated current
$J^\mu\rightarrow J^\mu_{\rm  IA}+
\lambda {J}^\mu_{\rm ex}$ where the interaction current and the
interaction satisfy  the
Ward-Takahashi identity (\ref{Ward2}). [In this section we will again
limit the discussion to nonidentical particles.]~ Here the parameter
$\lambda$ has  been  introduced to
assist in the counting of occurrences of the interaction ${V}$
and the  associated exchange current $J^\mu_{\rm ex}$ and will
eventually be set to unity in all calculations. From
(\ref{quasipotential}) it is clear that the  quasipotential can be
written as a series in $\lambda$ as
\begin{equation}
{U}=\sum_{N=1}^\infty \lambda^N{U}^{(N)}
\label{urankn}
\end{equation}
Substituting this into (\ref{quasipotential}) and collecting the
coefficients  of the various powers of $\lambda$, we can identify
the quasipotential of the $N$-th  rank ${U}^{(N)}$ as
\begin{eqnarray}
{U}^{(1)} &=& {V} \, , \label{quasiu0}\\
{U}^{(N)} &=& - {V} \Delta g_1  {U}^{(N-1)} \, ,
  \quad N> 1\, .
\label{quasiuN}
\end{eqnarray}
Similarly, using (\ref{Current1}) implies that the effective
current can also be expanded
\begin{equation}
J^{\mu}_{11}=\sum_{N=1}^\infty \lambda^N J^{(N)\mu}_{11}
\end{equation}
where the $N$-th rank contributions to the effective current are
$J^{(N)\mu}_{11} = J^{(N)\mu}_{\rm IA,eff}+ J^{(N)\mu}_{\rm ex,eff}$.
For $J^{(N)\mu}_{\rm IA,eff}$ these contributions are
\widetext
\begin{eqnarray}
  J^{(0)\mu}_{\rm IA,eff} &=& {\cal Q}_1 J^\mu_{\rm IA} {\cal Q}_1 \, ,
\label{jeffia0}\\
  J^{(1)\mu}_{\rm IA,eff} &=& -{\cal Q}_1 \left({U}^{(1)} \Delta g_1
J^\mu_{\rm IA} +
  J^\mu_{\rm IA} \Delta g_1 {U}^{(1)} \right) {\cal Q}_1   \, ,
\label{jeffia1}\\
  J^{(N)\mu}_{\rm IA,eff} &=& - {\cal Q}_1 \left({U}^{(N)} \Delta g_1
J^\mu_{IA} +
  J^\mu_{IA} \Delta g_1 {U}^{(N)} -
 \sum^{N-1}_{M=1} {U}^{(N-M)} \Delta g_1 J^\mu_{IA}
 \Delta g_1 {U}^{(M)} \right) {\cal Q}_1\, ,\nonumber\\
&&\qquad\qquad\qquad\qquad\qquad\qquad\qquad\qquad\qquad
\qquad\qquad\qquad {\rm if} \quad N>1\, , \label{jeffia}
\end{eqnarray}
and
\begin{eqnarray}
  J^{(0)\mu}_{\rm ex,eff}&=&0 , 
\label{jeffex0}\\
 J^{(1)\mu}_{\rm ex,eff}&=&{\cal Q}_1 J^\mu_{\rm ex} {\cal Q}_1  ,
 \label{jeffex1}\\
 J^{(2)\mu}_{\rm ex,eff} &=& - {\cal Q}_1 \left( {U}^{(1)} \Delta g_1
J^\mu_{\rm ex} +
  J^\mu_{ex} \Delta g_1 {U}^{(1)} \right){\cal Q}_1  , 
\label{jefx2}\\
 J^{(N)\mu}_{\rm ex,eff} &=& - {\cal Q}_1 \left({U}^{(N-1)} \Delta
g_1 J^\mu_{\rm ex} +
  J^\mu_{ex} \Delta g_1 {U}^{(N-1)} -
 \sum^{N-2}_{M=1} {U}^{(N-M-1)} \Delta g_1 J^\mu_{\rm ex}
 \Delta g_1 {U}^{(M)} \right){\cal Q}_1 \, , \nonumber\\
&&\qquad\qquad\qquad\qquad\qquad\qquad\qquad\qquad\qquad
\qquad\qquad\qquad {\rm if}\quad N>2\, .
\label{jeffex}
\end{eqnarray}
\narrowtext
At the lowest rank $N=0$ the particles do not
interact and only disconnected diagrams [which are not fully described
by the current (\ref{Current1})] occur. To get a nontrivial
description of interacting particles and their effective currents one
has to include at least the rank $N=1$ terms.

It is easy to show that a theory truncated at rank $N$ is gauge invariant
(and also  covariant of course) provided all terms up to and including rank
$N$ are included.  To do this we use Eqs.~(\ref{quasiu0})--(\ref{quasiuN})
and rules (\ref{rule1})--(\ref{rule5}) in Appendix B to show that
\begin{eqnarray}
q_\mu\left(J^{(0)\mu}_{11} + J^{(1)\mu}_{11}\right)
  =&& Q_1\left[ e_2(q),G_2^{-1} + {U}^{(1)}_{11}\right] Q_1
\label{WTspecNa}\\
q_\mu J^{(N)\mu}_{11}
  =&& Q_1\left[ e_2(q),{U}^{(N)}_{11}\right] Q_1
\label{WTspecN}
\end{eqnarray}
where (\ref{WTspecNa}) holds for the sum of $N=0$ and $N=1$ terms, and
(\ref{WTspecN}) for any finite $N \ge 2$.  Hence all terms linear
in $e_1$ cancel exactly in the truncated WT identities,
(\ref{WTspecNa}) and (\ref{WTspecN}), just as they do  in the
untruncated identity, Eq.~(\ref{WTspec}), and the results
(\ref{WTspecNa}) and (\ref{WTspecN}) are completely consistent with
(\ref{WTspec}).  We have shown that an effective current which is the
sum of terms up to any rank $N_{\rm max}\ge 1$ is gauge invariant
provided only that the quasipotential and the current include all
contributions up to rank $N_{\rm max}$.  Furthermore, the derivation
required only that the  BS kernel and BS current satisfy (\ref{Ward2});
they are otherwise unspecified.

Now consider a Bethe-Salpeter potential consisting of two independent
contributions
\begin{equation}
{V} = \lambda_1 \, {V}_1 +   \lambda_2 \, {V}_2 \, .
\label{vdecom2}
\end{equation}
with corresponding exchange currents $\lambda_1
J^\mu_{1,\rm ex}+ \lambda_2 J^\mu_{2,\rm ex}$ where the two components of this
current satisfy (\ref{Ward2}) with the corresponding components of the
potential. Examination of (\ref{quasipotential}) indicates that the
quasipotential can be expanded in the form
\begin{equation}
{U}=\sum_{N_1,N_2=0}^\infty \lambda_1^{N_1}\lambda_2^{N_2}
{U}^{(N_1,N_2)}\, ,
\end{equation}
where Eq.~(\ref{quasipotential}) gives
\widetext
\begin{eqnarray}
 &&{U}^{(0,0)}=0 \, , \\
 &&{U}^{(1,0)}={V}_1 \, ,\\
 &&{U}^{(0,1)}={V}_2\, ,\\
 &&{U}^{(N_1,0)}=-{V}_1\Delta g_1{U}^{(N_1-1,\,0)}\, ,
 \quad N_1>1 \, , \\
 &&{U}^{(0,N_2)}=-{V}_2\Delta g_1{U}^{(0,\,N_2-1)}\, ,
 \quad N_2>1 \, ,\\
 &&{U}^{(N_1,N_2)}=-{V}_1\Delta g_1{U}^{(N_1-1,\,N_2)}
  -{V}_2\Delta g_1{U}^{(N_1,\,N_2-1)}\, , \quad N_1,N_2\ge 1\, .
\end{eqnarray}
Using (\ref{GrossCurrent1}), the corresponding contributions to the
effective  current are
\begin{eqnarray}
&&J^{(0,0)\mu}_{\rm IA,eff}={\cal Q}_1J^\mu_{\rm IA}{\cal Q}_1\, , \\
&&J^{(N_1,N_2)\mu}_{\rm IA,eff}=-{\cal Q}_1\biggl\{{U}^{(N_1,N_2)}\Delta g_1
J^\mu_{\rm  IA}
  +J^\mu_{\rm IA} \Delta g_1 {U}^{(N_1,N_2)} \nonumber\\
 &&\phantom{J^{(N_1,N_2)\mu}_{\rm IA,eff}=} -\sum_{M_1=0}^{N_1}
\sum_{M_2=0}^{N_2} {U}^{(N_1-M_1,\, N_2-M_2)}
   \Delta g_1 J^\mu_{\rm IA} \Delta g_1  {U}^{(M_1,M_2)}
\biggr\}{\cal Q}_1 \, ,  \quad    N_1 {\rm or}\; N_2>1\, , \\
&&J^{(0,0)\mu}_{\rm ex,eff}=0\, , \\
&&J^{(1,0)\mu}_{\rm ex,eff}={\cal Q}_1J^\mu_{1,\rm ex} {\cal Q}_1\, , \\
&&J^{(0,1)\mu}_{\rm ex,eff}={\cal Q}_1 J^\mu_{2,\rm ex} {\cal Q}_1\, , \\
&&J^{(N_1,N_2)\mu}_{\rm ex,eff}=
   -{\cal Q}_1\Biggl\{{U}^{(N_1-1,\,N_2)}\Delta g_1 J^\mu_{1,\rm ex}
   +{U}^{(N_1,\,N_2-1)}\Delta g_1 J^\mu_{2,\rm ex} \nonumber\\
   &&\phantom{J^{(N_1,N_2)\mu}_{\rm ex,eff}=}
+ J^\mu_{1,\rm ex}\,\Delta g_1{U}^{(N_1-1,\,N_2)}
   + J^\mu_{2,\rm ex}\,\Delta g_1{U}^{(N_1,\,N_2-1)} \nonumber\\
   &&\phantom{J^{(N_1,N_2)\mu}_{\rm ex,eff}=}
-\sum_{M_1=0}^{N_1-1}\sum_{M_2=0}^{N_2}
  {U}^{(N_1-M_1-1,\,N_2-M_2)}
   \Delta g_1 J^\mu_{1,\rm ex}\,\Delta g_1
{U}^{(M_1,M_2)} \nonumber\\
&&\phantom{J^{(N_1,N_2)\mu}_{\rm ex,eff}=}
-\sum_{M_1=0}^{N_1}\sum_{M_2=0}^{N_2-1}
{U}^{(N_1-M_1,\,N_2-M_2-1)}
   \Delta g_1 J^\mu_{2,\rm ex}\,\Delta g_1 {U}^{(M_1,M_2)}\Biggr\}{\cal Q}_1
\, ,   \quad  N_1,N_2>1 \, .
\label{j2effex}
\end{eqnarray}
The divergence of the effective current is then
\begin{eqnarray}
q_\mu J^{(0,0)\mu}_{11}
  =&& Q_1\left[ e_2(q), G_2^{-1}\right] Q_1 \, , \nonumber\\
q_\mu J^{(N_1,N_2)\mu}_{11}
  =&& Q_1\left[ e_2(q), {U}^{(N_1,N_2)}_{11}\right]Q_1 \quad N_1\;
 {\rm or}\; N_2>1 \, .
\end{eqnarray}
\narrowtext
This implies that if all terms up to $N_{1\rm max}$ and $N_{2\rm max}$
are  retained in the quasipotential and the effective current that the
Ward-Takahashi identity will be satisfied. Note that $N_{1\rm max}$ and
$N_{2\rm max}$ do not have to be equal. That is, contributions from the
two  parts of the interaction can be truncated at different orders
without  disturbing the Ward-Takahashi identity.

The implication of these two results is that it is possible to truncate
the  quasipotential and interaction current in a consistent fashion
without  disturbing the Ward-Takahashi identities and that the truncation
can happen at  arbitrary orders. Indeed, from this it is clear that the
requirement of current  conservation places little constraint on the
truncation of the equation. Some  other physical consideration must
then determine the method of truncation of  these quantities.

An often used approximation to the Bethe-Salpeter equation is to collect
contributions to the kernel containing the same number of boson exchanges. This
is a natural procedure in the case of a perturbative approximation for a weak
coupled field theory. This approximation is also used in relativistic models of
the nucleon-nucleon system where the justification is that irreducible
contributions with increasing numbers of exchanged bosons have a
shorter range and tend to have a  small effect on the wave functions
and low energy scattering amplitudes.

Consider an interaction following from multiple
exchanges of the single type of boson
\begin{equation}
 {V} = \sum_{n=1}^\infty   {V}^{(n)}   \, ,
\label{vdecom}
\end{equation}
where the superscript $n$ denotes the number of exchanged bosons and
${V}^{(n)}$ is an irreducible contribution to the Bethe-Salpeter
kernel. Again, from (\ref{Ward2}) it follows that the
Bethe-Salpeter exchange currents can be decomposed in a similar way
\begin{equation}
 J^{\mu}_{\rm ex} = \sum_{n=1}^\infty  J^{(n)\mu}_{\rm ex} \, ,
\end{equation}
and the Ward-Takahashi identity is satisfied separately for each $n$.
Actually, in passing to our quasipotential framework we can formally
consider each set of Bethe-Salpeter-irreducible contributions of fixed
$n$ to  be independent contributions in the sense considered in the
second case  discussed above. The quasipotential and effective current
for each contribution  could then be truncated independently of the
others.

However, it has been shown that the convergence of the Gross
equation is  improved, in some cases, by a delicate cancellation of
crossed-box diagrams and subtracted box  diagrams of the same order in $n$
arising from the iteration of the  quasipotential equation. Therefore, the
physical consideration of convergence may  require that contributions to
the quasipotential with a fixed number of boson  exchanges also be
collected together. That is, the quasipotential can also be  expanded
\begin{equation}
 {U} = \sum_{n=1}^\infty   {U}^{(n)}   \, ,
\end{equation}
where $n$ is the number of exchanged bosons contributing to
${U}^{(n)}$. Substituting this into (\ref{quasipotential}) gives
\begin{eqnarray}
{U}^{(1)}&=&{V}^{(1)} , \\
{U}^{(n)}&=&{V}^{(n)}-\sum_{a=1}^{n-1}{V}^{(n-a)}
\Delta g_1 {U}^{(a)} , \quad n>1 .
\end{eqnarray}
Using (\ref{GrossCurrent1}), the corresponding contributions to the
effective current are
\widetext
\begin{eqnarray}
J^{(0)\mu}_{\rm IA,eff}&=&{\cal Q}_1J^\mu_{\rm IA}{\cal Q}_1 \, , \\
J^{(1)\mu}_{\rm IA,eff}&=&-{\cal Q}_1\left\{ {U}^{(1)}\Delta g_1
			J^\mu_{\rm IA}
                          +J^\mu_{\rm IA}\Delta g_1 {U}^{(1)}\right\}
                           {\cal Q}_1 \, , \\
J^{(n)\mu}_{\rm IA,eff}&=&-{\cal Q}_1\left\{ {U}^{(n)}\Delta g_1
J^\mu_{\rm IA}
                          +J^\mu_{\rm IA}\Delta g_1 {U}^{(n)} -
    \sum_{a=1}^{n-1}{U}^{(n-a)}\Delta g_1J^\mu_{\rm IA} \Delta g_1
    {U}^{(a)} \right\} {\cal Q}_1 \, , \quad n>1\, , \\
J^{(0)\mu}_{\rm ex,eff}&=&0 \, , \\
J^{(1)\mu}_{\rm ex,eff}&=&{\cal Q}_1J^{(1)\mu}_{\rm ex}{\cal Q}_1\, , \\
J^{(2)\mu}_{\rm ex,eff}&=&{\cal Q}_1\left\{ J^{(2)\mu}_{\rm ex}
- {U}^{(1)}\Delta g_1 J^{(1)\mu}_{\rm ex}
                          -J^{(1)\mu}_{\rm ex}\Delta g_1
{U}^{(1)}\right\} {\cal Q}_1\,  ,
\label{jneff2}\\
J^{(n)\mu}_{\rm ex,eff}&=&{\cal Q}_1 \Biggl\{J^{(n)\mu}_{\rm ex}
-\sum_{a=1}^{n-1} {U}^{(n-a)}\Delta g_1 J^{(a)\mu}_{\rm ex}
- \sum_{a=1}^{n-1}J^{(n-a)\mu}_{\rm ex} \Delta g_1 {U}^{(a)}
\nonumber\\
&&\phantom{{\cal Q}_1
\Biggl\{}+\sum_{a=1}^{n-2}\sum_{b=1}^{n-a-1}{U}^{(n-a-b)}
\Delta g_1 J^{(a)\mu}_{\rm ex} \Delta g_1 {U}^{(b)}\Biggr\}{\cal Q}_1
\, , \quad n>2 \, .
\label{jneff}
\end{eqnarray}
\narrowtext
The divergence of this effective current is (see Appendix B)
\begin{eqnarray}
  q_\mu J^{(0)\mu}_{11} &=& Q_1 \left[e_2(q), G_2^{-1}\right]\, Q_1 \\
  q_\mu J^{(n)\mu}_{11}  &=&
   \left[ e_2(q), {U}^{(n)}_{11}\right] \quad n\ge 1  \, .
\label{WTnex}
\end{eqnarray}
This implies that the Ward-Takahashi identity is satisfied if the
quasipotential and effective current  include  all contributions from
boson exchanges up to some $n_{\rm max}$. This can be easily generalized
to include additional  kinds of bosons. From the second case presented
above it is also clear that  the equations can be truncated at different
numbers of boson exchanges for each  type of boson. For example, a meson
exchange model of the nucleon-nucleon  interaction could contain
contributions from up to two pion exchanges, but  heavier meson
contributions could be truncated at the one-boson-exchange level.

\section{Conclusions}

This paper develops a detailed algebraic treatment of the  spectator
or Gross description of strongly interacting two-particle systems
in the presence of an external electromagnetic field (treated to first
order). Our factorization of the five-point function follows naturally
from the original definition of the spectator equations.

We start from the Bethe-Salpeter formulation, i.e., we assume that
the underlying dynamics  is known in principle and that
it generates a series of Feynman diagrams which specifies both the
interactions of two-nucleon system (Bethe-Salpeter equation) and the
interaction of the two-nucleon system with an external electromagnetic
field (Bethe-Salpeter exchange currents). The Bethe-Salpeter currents
satisfy a  Ward-Takahashi identity involving the Bethe-Salpeter four-point
propagator.

The spectator description is shown to result from rearranging these
sets of diagrams, expressing the dynamics effectively in terms
of a modified free two-nucleon propagator: in intermediate states
one of the nucleons is restricted to its positive energy mass shell.
The parts of the original diagrams in which this constraint does not
hold are summed  into a new effective interaction kernel
(quasipotential) and an effective current (interaction current). The
effective current satisfies a  Ward-Takahashi identity with the
corresponding four-point spectator propagator, so that the current is
conserved. When all terms are included, the wave functions and current
matrix elements are identical to those of the Bethe-Salpeter formalism.

In applications, the whole infinite set of diagrams is not generally
included, and we show that the series can be truncated to any finite order
and still preserve gauge invariance.  Most applications of the Gross
formalism have been made using the lowest (second-order)
one-meson-exchange approximation. Formally, this paper defines a
consistent formulation for any finite order, and also shows
that it is possible, for example,  to include consistently the
forth-order two-meson exchange contributions for some of the more
important mesons (perhaps only the pion) while at the same time
limiting the treatment of heavier mesons to the lowest, second-order.

Although we have confined the arguments of this paper to the
construction of  electromagnetic current matrix elements, the method is
general and can be used, for example, to treat weak and axial vector
currents.  The extension of this formalism to three-particle systems
will be presented in a future paper \cite{avg98}.

\acknowledgments

We are happy to acknowledge the support of the DOE through
Jefferson Laboratory, and  we gratefully acknowledge the
support of the DOE through grant No.~DE-FG02-97ER41032 (for FG)
and DE-FG02-97ER41028 (for JWVO).

\renewcommand{\theequation}{A.\arabic{equation}}
\setcounter{equation}{0}
\section*{Appendix A: The one-body current for particle 1}

In this Appendix we briefly discuss the comments of Kvinikhidze and
Blankleider \cite{kb97} in more detail.

To illustrate the issue, consider the following contour integral
\begin{equation}
{\cal I}={1\over2\pi i}\int_C {f(z)\, dz  \over
(z-z_1-i\epsilon_1)(z-z_2-i\epsilon_2)}
\, , \label{a1}
\end{equation}
where the contour $C$ encloses the two poles at $z_1$ and $z_2$,
$f(z)$ is analytic inside of the contour, and the {\it limit
$\epsilon_i\to0$ is implied\/}.  Evaluation of the integral is
straightforward, and gives
\begin{eqnarray}
{\cal I}=&&{1 \over z_1-z_2+i\,\delta\epsilon}
\Bigl\{f(z_1+i\epsilon_1) -  f(z_2+i\epsilon_2)\Bigr\}\nonumber\\
\rightarrow&&  \, f'(z_1+i\epsilon_1) \to  \,   f'(z_1)\qquad 
{\rm as}\; z_1 \to z_2 \, ,
\end{eqnarray}
where the contour $C$ encloses the two positive energy
poles only, $\delta\epsilon=\epsilon_1-\epsilon_2$.
Note that zero in the denominator at
$z_1-z_2+i\,\delta\epsilon=0$ is canceled exactly by a zero in the
numerator, so the final result has no singularity.

In the derivation of the one-body current for particle one, leading to
Eq.~(\ref{gjg}),  we are confronted with a similar integral.  In that
case, in the Breit frame, the integral comparable to (\ref{a1}) is
\widetext
\begin{eqnarray}
{\cal J}^\mu_1=&&\int_C {dk_0\over2\pi i}\,  {J_1^\mu(k+{\textstyle
{1\over2}}q,k-{\textstyle {1\over2}}q)\over
(E_++k_0-i\epsilon_1)(E_-+k_0-i\epsilon_2)
(E_+-k_0-i\epsilon_1)(E_--k_0-i\epsilon_2)}\nonumber\\
=&&{1\over (E_--i\epsilon_2)^2-(E_+-i\epsilon_2)^2} \left\{{J_1^\mu({k}_+
+{\textstyle {1\over2}}q,k_+ -{\textstyle {1\over2}}q)\over
2(E_+ -i\epsilon_1)} -
{J_1^\mu({k}_- +{\textstyle {1\over2}}q,k_- -{\textstyle {1\over2}}q)\over
2(E_- -i\epsilon_2) } \right\}
\, , \label{xx}
\end{eqnarray}
where the contour $C$ encloses the two positive energy
poles only, $E_\pm=\sqrt{m^2+\left({\bf k}\pm {\scriptstyle {1\over2}}{\bf
q}\right)^2}$, $k_+=(E_+-i\epsilon_1, {\bf k})$, and
$k_-=(E_--i\epsilon_2, {\bf k})$.  Once again, the zero in the
denominator  at $E_- -E_+ +i\,\delta\epsilon=0$ is canceled exactly by a
zero in the numerator, so the final result has no singularity. However,
the first two terms in the last line of Eq.~(\ref{gjg}) (identical to the
last two terms of Eq.~(2.33) in Ref.~\cite{norm}), in the notation of
Eq.~(\ref{xx}), become
\begin{equation}
{\cal Q}_1\,J^\mu_1\, G_1+
G_1\,J^\mu_1\,{\cal Q}_1 \simeq \left\{
{J_1^\mu({k}_+ +{\textstyle {1\over2}}q,k_+ -{\textstyle {1\over2}}q)\over
2E_+ (E_-^2-E_+^2-i\epsilon)} + {J_1^\mu({k}_-
+{\textstyle {1\over2}}q,k_- -{\textstyle {1\over2}}q)\over
2E_- (E_+^2-E_-^2-i\epsilon) } \right\} \, , \label{yy}
\end{equation}
\narrowtext
where we have retained the $i\epsilon$ terms in the $G_1$ propagators.
As Eq.~(\ref{xx}) shows, these are not the correct $i\epsilon$ factors,
and they should be dropped immediately by taking the
$\epsilon\to0$ limit (as we are instructed to do).  Dropping them gives a
result identical to (\ref{xx}).  Were we to (incorrectly) retain the
$i\epsilon$'s in Eq.~(\ref{yy}) and expand the denominators into a
principal value term and a delta function, the resulting delta function
contributions to Eq.~(\ref{yy}) would {\it not\/} cancel, giving an
incorrect contribution to the current of the type ${\cal Q}_1
\,J^\mu_1\,{\cal Q}_1$.  We agree with  Kvinikhidze and
Blankleider that this contribution is spurious.  It has not
been included in any previous applications \cite{GR87}--\cite{deutff},
and is eliminated by taking the $\epsilon\to0$ limit [or simply dropping the
$i\epsilon$ terms from Eq.~(\ref{yy})] after the contour integration has
been carried out.

\renewcommand{\theequation}{B.\arabic{equation}}
\setcounter{equation}{0}
\section*{Appendix B: Gauge invariance for truncated currents.}

In this appendix we verify the gauge invariance of the truncated currents
introduced in Section IV. 

First, for the purpose of further discussion
it is convenient to split the divergence of the total untruncated current
(\ref{WTspec}) into two parts corresponding to the divergences of the
effective currents $J^\mu_{\rm IA,eff}$ and $J^\mu_{\rm ex,eff}$, introduced
in (\ref{GrossCurrent1}) and generated by 
the one-particle $J^\mu_{\rm IA}$ and the interaction $J^\mu_{\rm ex}$ 
Bethe-Salpeter currents, respectively. In particular
\widetext
\begin{eqnarray}
 q_\mu J^\mu_{\rm IA,eff} &=& {\cal Q}_1 \biggl( [e_2(q), \, G_2^{-1} ] -
  [e_1(q),\, U]- U\, [e_1(q)+ e_2(q), \, \Delta g_1]\, U \biggr) {\cal Q}_1
   \, , \label{divia}     \\
  q_\mu J^\mu_{\rm ex,eff}&=& {\cal Q}_1 \biggl(  [e_1(q)+ e_2(q), \, U] 
  + U\, [e_1(q)+ e_2(q), \, \Delta g_1]\, U \biggr) {\cal Q}_1 \, . 
\label{divex}
\end{eqnarray}
The relation (\ref{divia}) follow from identities (\ref{rule1}--\ref{rule5}) 
and its derivation can be repeated without
any modification  for the corresponding truncated effective currents.
In deriving (\ref{divex}) one has to use 
the quasipotential equation (\ref{quasipotential}) and
more care is needed to get the divergence for the 
 truncated $J^\mu_{\rm ex,eff}$.

Let us now consider the truncation by the rank $N$ of the quasipotential
$U^{(N)}$, as defined in eqs.\ (\ref{urankn}--\ref{jeffex}). 
Using again the identities (\ref{rule1}--\ref{rule5}) one gets
\begin{eqnarray}
 q_\mu \biggl( J^{(0)\mu}_{\rm IA,eff}+ J^{(1)\mu}_{\rm IA,eff}
 + J^{(1)\mu}_{\rm ex,eff} \biggr) &= & {\cal Q}_1 
 \biggl( [e_2(q), \, G_2^{-1} ]-
 [e_1(q),\, V] + [e_1(q)+ e_2(q), \, V] \biggr) {\cal Q}_1  \nonumber\\
 &=& {\cal Q}_1   \biggl( [e_2(q), \, G_2^{-1} + U^{(1)} ]\, 
 \biggr) {\cal Q}_1 \, ,
\end{eqnarray}
and repeating the derivation of (\ref{divia}) for truncated 
quasipotential with $N > 1$  
\begin{equation}
 q_\mu J^{(N)\mu}_{\rm IA,eff} = {\cal Q}_1 \biggl(   -
  [e_1(q),\, U^{(N)} ]- \sum_{M=1}^{N-1}
  U^{(N-M)}\, [e_1(q)+ e_2(q), \, \Delta g_1]\, U^{(M)} \, 
  \biggr) {\cal Q}_1 \, . 
\label{diviaN}
\end{equation}   
For the corresponding $J^{(N)\mu}_{\rm ex,eff}$ as given by 
(\ref{jefx2}--\ref{jeffex}) we get
\begin{eqnarray}
q_\mu J^{(N)\mu}_{\rm ex,eff} &=& {\cal Q}_1 \biggl( 
 U^{(N-1)} \Delta g_1 V e(q)
 - U^{(N-1)} \Delta g_1 e(q) V -  e(q) V \Delta g_1 U^{(N-1)} \nonumber\\
&& + V  e(q) \Delta g_1 U^{(N-1)}  + \sum_{M=1}^{N-2} 
 U^{(N-M-1)} \Delta g_1 [e(q), \, V ]\, \Delta g_1 U^{(M)} 
 \, \biggr) {\cal Q}_1 
 \nonumber\\
&=& {\cal Q}_1 \biggl( [e(q), \, U] + V e(q) \Delta g_1 U^{(N-1)}
+ \sum_{M=1}^{N-2} U^{(N-M)} e(q) \Delta g_1 U^{(M)} \nonumber\\
&& - U^{(N-1)} \Delta g_1 e(q) V 
- \sum_{M=1}^{N-2} U^{(N-M-1)}  \Delta g_1  e(q) U^{(M+1)}\, 
\biggr) {\cal Q}_1  \nonumber\\
&=& {\cal Q}_1 \biggl( [e_1(q)+ e_2(q), \, U] +
 \sum_{M=1}^{N-1} U^{(N-M)} [e_1(q)+ e_2(q), \Delta g_1 ] U^{(M)} \, 
 \biggr) {\cal Q}_1  \, ,
\label{divexN}
\end{eqnarray}
\narrowtext
where we introduced  the shorthand notation $e(q)= e_1(q)+ e_2(q)$ 
in intermediate steps. The derivation is valid for $N>1$, though
for $N=2$ some summations are empty. Clearly, the sum of (\ref{diviaN})
and (\ref{divexN}) gives (\ref{WTspecN}). 

This derivation can be repeated  for the case of two interactions
defined by Eqs.\ (\ref{vdecom2}--\ref{j2effex}). In this case one has to
inspect the bounds of the summations carefully when the quasipotential equation
is used, since the summations contain $U^{(0,0)}=0$ and therefore 
terms like $V_1 \Delta g_1 U^{(M_1,M_2)}$ should be treated separately  
if $M_1=M_2=0$.

Finally, for the truncation by the number of exchanged mesons, as defined
by Eqs.\ (\ref{vdecom}--\ref{jneff}), we obtained exactly as before
\widetext
\begin{eqnarray}
  q_\mu J^{(0)\mu}_{\rm IA,eff} &= &
  {\cal Q}_1 [e_2(q), \, G_2^{-1} ]  {\cal Q}_1 \, , \\
  q_\mu J^{(n)\mu}_{\rm IA,eff} &= & {\cal Q}_1 \biggl(   -
  [e_1(q),\, U^{(n)} ]- \sum_{a=1}^{n-1}
  U^{(n-a)}\, [e_1(q)+ e_2(q), \, \Delta g_1 ]\, U^{(a)} \, 
  \biggr) {\cal Q}_1 \, , 
\label{diviann}
\end{eqnarray}
where (\ref{diviann}) is valid for $n >0$ and the sum does not contribute
for $n=1$. Both currents $J^{(n)\mu}_{\rm ex,eff}$ from
(\ref{jneff2}) and (\ref{jneff}) can be considered at the
same time and we get for $n>0$ the divergence 
\begin{eqnarray}
  q_\mu J^{(n)\mu}_{\rm ex,eff} &= & {\cal Q}_1  \biggl( 
  e(q)\, 
  \left[ V^{(n)} - \sum_{a=1}^{n-1}  V^{(n-a)} \Delta g_1 U^{(a)} \right]
- \left[ V^{(n)} - \sum_{a=1}^{n-1} U^{(n-a)} \Delta g_1 V^{(a)} \right]\,
  e(q)
\nonumber\\
&& + \sum_{a=1}^{n-1}  V^{(n-a)} e(q) \Delta g_1 U^{(a)} -  \sum_{b=1}^{n-2}
 \sum_{a=1}^{n-b-1}  U^{(n-a-b)} \Delta g_1 V^{(b)} e(q) \Delta g_1 U^{(a)}
 \nonumber\\
&& - \sum_{a=1}^{n-1}  U^{(n-a)} \Delta g_1 e(q) V^{(a)} +  \sum_{b=1}^{n-2}
  \sum_{a=1}^{n-b-1} U^{(n-a-b)} \Delta g_1 e(q) V^{(b)}  \Delta g_1 U^{(a)}
 \, \biggr) {\cal Q}_1  \nonumber\\
&=& {\cal Q}_1 \biggl(  [e(q), \, U^{(n)}] \nonumber\\
&&+ \sum_{a=1}^{n-1}  V^{(n-a)} e(q) \Delta g_1 U^{(a)} -  \sum_{a=1}^{n-2}
\left[ \sum_{b=1}^{n-a-1} U^{(n-a-b)} \Delta g_1 V^{(b)} \right] 
 e(q) \Delta g_1 U^{(a)} \nonumber\\
&& - \sum_{a=1}^{n-1}  U^{(n-a)} \Delta g_1 e(q) V^{(a)}+ \sum_{b=1}^{n-2}
 \sum_{c=b+1}^{n-1}  U^{(n-c)} \Delta g_1  e(q) V^{(b)} \Delta g_1 U^{(c-b)}
 \, \biggr) {\cal Q}_1 \nonumber\\
&=&  {\cal Q}_1 \biggl(  [e(q), \, U^{(n)}] + 
 \sum_{a=1}^{n-1} U^{(n-a)} e(q) \Delta g_1 U^{(a)} \nonumber\\
&& - \sum_{a=1}^{n-1}  U^{(n-a)} \Delta g_1 e(q) V^{(a)}+ \sum_{c=2}^{n-1}
 U^{(n-c)} \Delta g_1  e(q) \sum_{b=1}^{c-1} V^{(b)} \Delta g_1 U^{(c-b)}
\, \biggr) {\cal Q}_1   \nonumber\\
&=&  {\cal Q}_1  \biggl(  [e(q), \, U^{(n)}] + 
\sum_{a=1}^{n-1} U^{(n-a)} [e(q), \, \Delta g_1] U^{(a)} \, 
\biggr) {\cal Q}_1  \, ,
\label{divexnn}
\end{eqnarray}
\narrowtext
where again some sums are empty for $n=1,2$. The sum of (\ref{diviann})
and (\ref{divexnn}) yields (\ref{WTnex}).

\end{document}